\documentclass{emulateapj}
\usepackage{natbib}
\usepackage{lscape}

\def\gsim{\;\rlap{\lower 2.5pt
 \hbox{$\sim$}}\raise 1.5pt\hbox{$>$}\;}
\def\lsim{\;\rlap{\lower 2.5pt
   \hbox{$\sim$}}\raise 1.5pt\hbox{$<$}\;}
\def\micron{~$\mu\textrm{m}$ }
\def\micronend{$\mu\textrm{m}$}
\def\microjy{~$\mu\textrm{Jy}$ }

\def\microjyend{~$\mu\textrm{Jy}$}

\begin{document}

\title{The AzTEC/SMA Interferometric Imaging Survey of Submillimeter-Selected High-Redshift Galaxies}
\author{Joshua D. Younger \altaffilmark{1,2,3,4}, Giovanni G. Fazio\altaffilmark{1}, Jia-Sheng Huang\altaffilmark{1}, Min S. Yun\altaffilmark{5}, Grant W. Wilson\altaffilmark{5}, Matthew L. N. Ashby\altaffilmark{1}, Mark A. Gurwell\altaffilmark{1}, Alison B. Peck\altaffilmark{6}, Glen R. Petitpas\altaffilmark{1}, David J. Wilner\altaffilmark{1}, David H. Hughes\altaffilmark{7}, Itziar Aretxaga\altaffilmark{7}, Sungeun Kim\altaffilmark{8}, Kimberly S. Scott\altaffilmark{5}, Jason Austermann\altaffilmark{5}, Thushara Perera\altaffilmark{5}, James D. Lowenthal\altaffilmark{9}}
\altaffiltext{1}{Harvard-Smithsonian Center for Astrophysics, 60 Garden Street, 
Cambridge, MA 02138}
\altaffiltext{2}{Current Address: Institute for Advanced Study, Einstein Drive, Princeton, NJ 08540}
\altaffiltext{3}{Hubble Fellow}
\altaffiltext{4}{jyounger@ias.edu}
\altaffiltext{5}{Astronomy Department, University of Massachusetts, Amherst, MA, 01003}
\altaffiltext{6}{Joint ALMA Office, El Golf 40, Las Condes, Santiago 7550108, Chile}
\altaffiltext{7}{Instituto Nacional de Astrof\'{i}sica, \'{O}ptica y Electr\'{o}nica (INAOE), Tonantzintla, Peubla, M\'{e}xico}
\altaffiltext{8}{Astronomy and Space Sciences Department, Sejong University, 98 Kwangjin-gu, Kunja-dong, Seoul, 143-747, Korea}
\altaffiltext{9} {Astronomy Department, Smith College,  Clark Science Center, Northampton, MA  01060} 

\begin{abstract}

We present results from a continuing interferometric survey of high-redshift submillimeter galaxies with the Submillimeter Array, including high-resolution (beam size $\sim 2$ arcsec) imaging of eight additional AzTEC 1.1mm selected sources in the COSMOS Field, for which we obtain six reliable (peak S/N $>5$ or peak S/N $>4$ with multiwavelength counterparts within the beam) and two moderate significance (peak S/N $>4$) detections.  When combined with previous detections, this yields an unbiased sample of millimeter-selected SMGs with complete interferometric followup.  With this sample in hand, we (1) empirically confirm the radio-submillimeter association, (2) examine the submillimeter morphology -- including the nature of submillimeter galaxies with multiple radio counterparts and constraints on the physical scale of the far infrared -- of the sample, and (3) find additional evidence for a population of extremely luminous, radio-dim submillimeter galaxies that peaks at higher redshift than previous, radio-selected samples.  In particular, the presence of such a population of high-redshift sources has important consequences for models of galaxy formation -- which struggle to account for such objects even under liberal assumptions -- and dust production models given the limited time since the Big Bang.  

\end{abstract}

\keywords{cosmology: observations -- galaxies: evolution -- galaxies: high-redshift -- galaxies: starburst -- galaxies: submillimeter -- galaxies: formation}

\section{Introduction}
\label{sec:intro}

Though they make up a very small fraction of the local infrared (IR) luminosity density, at $z\gsim 1$ IR-luminous galaxies \citep[LIRGs and ULIRGs:][]{sanders1996} become cosmologically important \citep[e.g.,][]{lefloch2005,magnelli2009} and contribute significantly to the diffuse extragalactic IR background \citep{hauser1998,kelsall1998,aredt1998,dwek1998,fixsen1998,pei1999,devlin2009}.  They are also thought to dominate the cosmic star formation rate (SFR) density at $z\gsim 1$ \citep{blain1999,blain2002,pascale2009}, drive the formation of luminous quasars \citep{sanders1988a,sanders1988b,hopkins2006,hopkins2007a,coppin2008.submmqso,ivison2008.agnsb,narayanan2009,narayanan2009b} and the most massive galaxies  \citep{scott2002,blain2004,swinbank2006,hopkins2007b,viero2009}.   As such, these objects provide powerful constraints on theoretical models \citep{baugh2005,swinbank2008}, and further study is essential to achieve a more thorough understanding of the birth and evolution of galaxies.

A significant population of high-redshift ULIRGs was first revealed at 850\micron by the Submillimeter Common User Bolometer Array \citep[SCUBA:][]{holland1999} at the James Clerk Maxwell Telescope \citep{smail1997,hughes1998,barger1998}.  At these wavelengths, the shape of the redshifted far-IR spectral energy distribution (SED) counteracts the effect of increasing luminosity distance to provide an unbiased view of dust-obscured star formation out to $z\sim 10$ \citep{blain1993}.  Since their initial discovery, observers have amassed extensive catalogs of these submillimeter galaxies \citep[SMGs: for a review, see][]{blain2002} in a number of fields across the sky at both submillimeter \citep[850\micronend:][]{eales1999,eales2000,cowie2002,scott2002,webb2003,serjeant2003,wang2004,coppin2006} and millimeter \citep[1100-1200\micronend:][]{ivison2004,greve2004,dannerbauer2004,laurent2005,dscott2006,bertoldi2007,scott2008,perera2008,wilson2008b,austermann2009} wavelengths.  

However, a more complete understanding of these objects has been hampered, in part, by the relatively poor resolution of submillimeter cameras (FWHM $\sim 10-18$ arcsec), which makes identification of multiwavelength counterparts inherently ambiguous. The first break-through came with very deep wide-field radio continuum surveys, which leveraged the local radio/far-IR correlation \citep[for a review, see][]{condon1992} in combination with statistical arguments to associate nearby radio sources with the submilliimeter emission \citep{ivison2002,ivison2007}.  This radio-submillimeter association permitted optical spectroscopic followup, which confirmed that SMGs lie at high-redshift \citep[median $z\approx 2.5$:][]{chapman2003a,chapman2005} and in turn enabled CO spectroscopic imaging \citep{neri2003,sheth2004,greve2005,tacconi2006,tacconi2008} which confirmed that most are young, gas-rich galaxies undergoing major mergers.   While the radio/far-IR correlation is broadly thought to apply at high-redshift \citep{garrett2002,gruppioni2003,appleton2004,boyle2007,younger2008.egsulirgs}, owing to the strong dimming of the radio continuum with increasing luminosity distance \citep[see e.g.,][]{carilli1999} existing radio-selected samples are biased towards somewhat lower redshift $1\lsim z \lsim 3$ objects.  While alternative counterpart identification techniques utilizing near and mid-IR imaging data exist \citep{ashby2006,pope2006,yun2008}, these too may be subject to biases which are difficult to quantify.

Since all the above mentioned techniques are inherently ambiguous, reliable counterpart identification remains one of the most challenging obstacles to detailed study of SMGs.  Though it currently requires a large investment of observing time, this motivates high-resolution interferometric imaging at the discovery wavelength utilizing facilities such as the Submillimeter Array \citep[SMA:][]{ho2004}, the Plateau de Bure Interferometer (PdBI), and the Caltech Millimeter Array (CARMA), which provide an order of magnitude improvement in the precision of absolute position measurements over single dish instruments and permit unambiguous identification of multiwavelength counterparts in higher resolution imaging data.  Previous interferometric observations at millimeter \citep{downes1999,frayer2000,dannerbauer2002,downes2003,genzel2003,kneib2005,greve2005,tacconi2006,tacconi2008} and submillimeter \citep{iono2006,younger2007,younger2008,younger2008highres,wang2007,dannerbauer2008,cowie2009} wavelengths have identified unambiguous counterparts for increasing numbers of radio-detected SMGs, and support the radio-submillimeter association.  In particular, several groups \citep[][M. Yun et al. in preparation]{younger2007,wang2007,wang2008,dannerbauer2008} found that the multiwavelength counterparts of radio-dim SMGs provided evidence for a significant population of SMGs at higher redshift than radio-selected samples; a result that has found recent support from spectroscopic observations of several individual objects \citep{capak2008,schinnerer2008,coppin2009,daddi2008,daddi2009b}.  

The existence of large numbers of $z\gsim 3$ ULIRGs has profound implications for models of galaxy formation and evolution.  Recent results from semi-analytic models \citep{baugh2005,swinbank2008} have successfully reproduced the observed SMG population at $z\approx 2.5$ -- though at the expense of a dramatic departure from a universal initial mass function (IMF) in starbursts.  However, even these tuned models struggle to produce a significant population of higher redshift SMGs \citep[see, e.g., Figure 4 and \S~4 in][]{coppin2009}.  Furthermore, given the limited time since the big bang, these high-redshift dusty starbursts provide constraints on models of dust production in the early universe.

In order to investigate the nature of these extreme objects, we have extended the original unbiased\footnote{In this context, we use the term `unbiased' to refer to a signal-to-noise limited sample of millimeter sources, selected entirely on the basis of their millimeter emission.  While this is not formally a flux-limited sample, given some sources with higher fitted fluxes but detections below the sigmal-to-noise limit, it is quite close to one -- especially at the high flux end.} survey \citep{younger2007} of 1.1mm selected SMGs with complete interferometric followup.  In this work, we present results of high-resolution SMA imaging at 890\micron for 8 new sources first identified in the AzTEC/COSMOS survey \citep{scott2008}.  

\section{Observations}
\label{sec:obs}

\begin{figure*}
\plotone{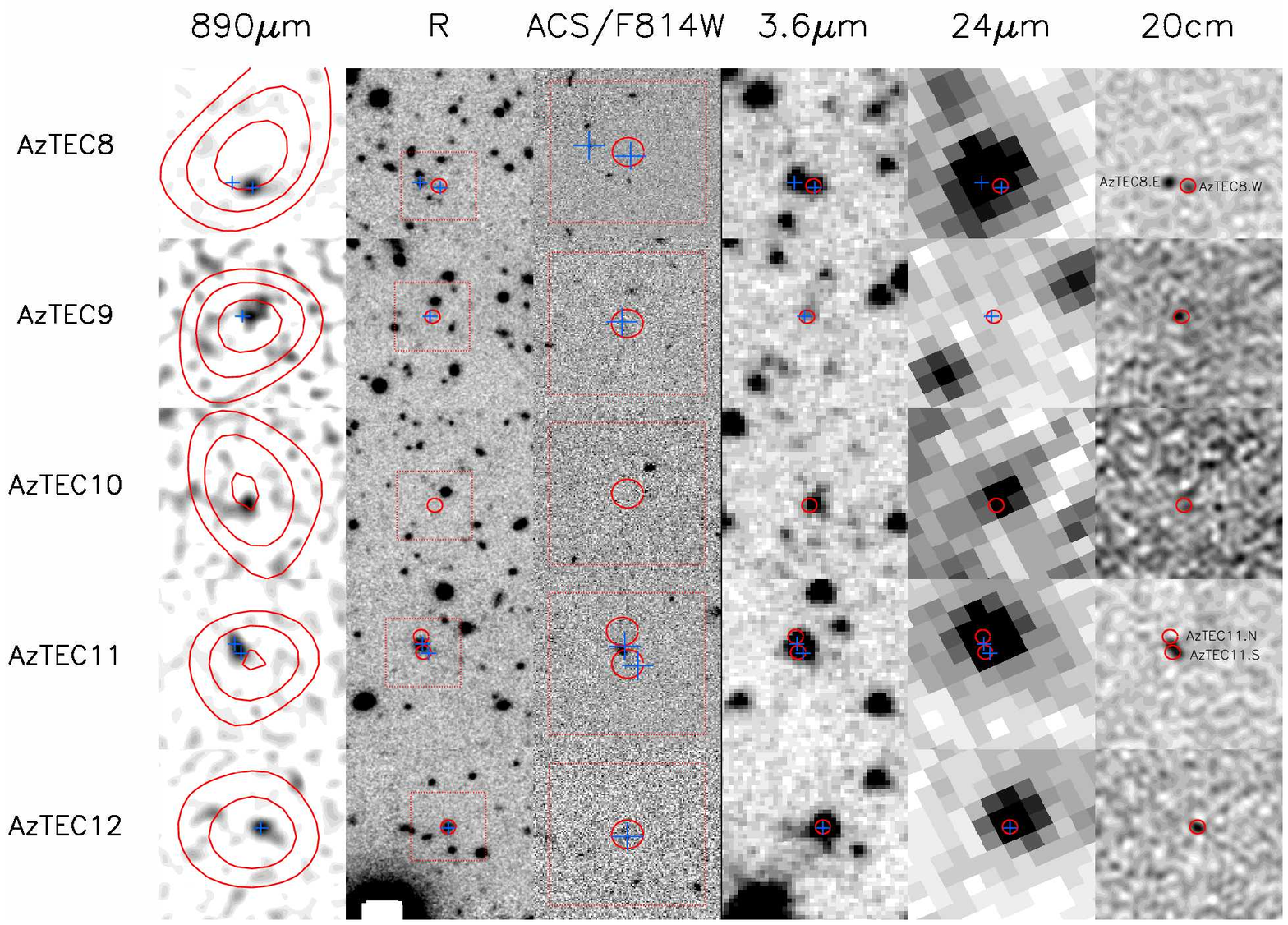}
\caption{Multiwavelength stamp images for the new objects in the sample, including (left to right) the SMA 890\micronend, Subaru R-band, ACS $z$-band imaging, IRAC 3.6\micron, MIPS 24\micron, and VLA 20cm data.  The red contours over the SMA data represent 3,4, $\ldots \times \sigma$ in the AzTEC 1100\micron map, the red circles over the remaining data indicate the SMA position and are 1 arcsec in diameter -- roughly 1/2 the SMA beam size, and blue crosses indicate the locations of radio sources within the AzTEC beam.  Each stamp is 25 arcsec on a side with the exception  of the ACS data, which shows a 5 arcsec box indicated by a dotted rectangle.  Multicomponent sources are labeled following Table~\ref{tab:photom}.}
\label{fig:stamps}
\end{figure*}

The COSMOS field \citep{scoville2007} benefits from an extraordinary wealth of deep, multi-wavelength coverage from the X-ray to the radio.  In this work, we utilize $i$ band imaging with the Advanced Camera for Surveys \citep[ACS:][]{ford1998} on board the {\it Hubble Space Telescope} to a depth of 27.2 magnitudes \citep[for point-sources at the $5\sigma$ level;][]{koekemoer2007}, a variety of ground-based optical and near-infrared imaging data \citep[see ][]{taniguchi2007,capak2007}, imaging by the Infrared Array Camera \citep[IRAC:][]{fazio2004} and the Multiband Imaging Photometer for Spitzer \citep[MIPS:][]{rieke2004} on board the {\it Spitzer Space Telescope} at 3.6, 4.5, 5.8, 8.0, and 24\micron to $5\sigma$ depths of $\sim$0.9, 1.7, 11.3, 14.6, and 71\microjy respectively \citep{sanders2007}, and 1.4 GHz radio continuum imaging to a mean rms depth of $\sim 10.5$\microjyend/beam with the Very Large Array \citep[VLA:][]{schinnerer2007}.  We also make use of photometric redshifts and stellar mass estimates from \citet{mobasher2006}.

The AzTEC/COSMOS survey covers 0.15 deg$^2$ of the COSMOS field at 1.1 mm with an rms noise level of 1.3 mJy/beam \citep{scott2008}.  The AzTEC/COSMOS catalog includes 50 sources with S/N $ \ge 3.5\sigma$, of which 10 sources have S/N $\ge5\sigma$.  For the SMA observations we chose the next eight highest significance sources ($4.5\lsim {\rm S/N} \lsim 5.5$) down from the original sample presented in \cite{younger2007}.  This yields, in total, an unbiased sample of 15 millimeter selected SMGs with complete interferometric followup.  Due to the high significance of the sources in our sample, the expected false detection is $\lsim$ 0.3 sources \citep[see Fig. 7 in][]{scott2008}.

The SMA observations were performed in the compact array configuration (beam size $\sim 2\arcsec$) at 345 GHz (full bandwidth 4 GHz from the combined sidebands) from December 2007 through March 2008.  The weather was generally excellent ($\tau_{\rm 225GHz} \lsim 0.08$), with typical rms noise levels of 1.0-1.5 mJy per track with $\sim 6$ hours of on-source integration.   The data were calibrated using the {\sc mir} software package \citep{scoville1993}, modified for the SMA.  Complex gain calibration was performed using the calibrator sources  J1058+015 ($\sim 3$ Jy, $\sim15^\circ$ away from targets) and J0854+201 ($\sim 1$ Jy, $\sim24^\circ$ away from targets).  Passband calibration was done using available strong calibrator sources, primarily 3C273 and Callisto. The absolute flux scale was set using observations of Callisto and is estimated to be accurate to better than 20\%.  Positions and fluxes of the COSMOS sources were derived from the calibrated visibilities using the {\sc miriad} software package \citep{sault1995}.  We also incorporated hourly integrations on a test quasar J1008+063 -- which is included in both the JVAS \citep{patnaik1992,browne1998} and VLBA Calibrator \citep{ma1998,beasley2002} catalogs of compact, flat-spectrum radio sources, and has an absolute position known to better than 20 mas -- to empirically verify the phase transfer.

\section{Results}
\label{sec:results}

\begin{figure*}
\plotone{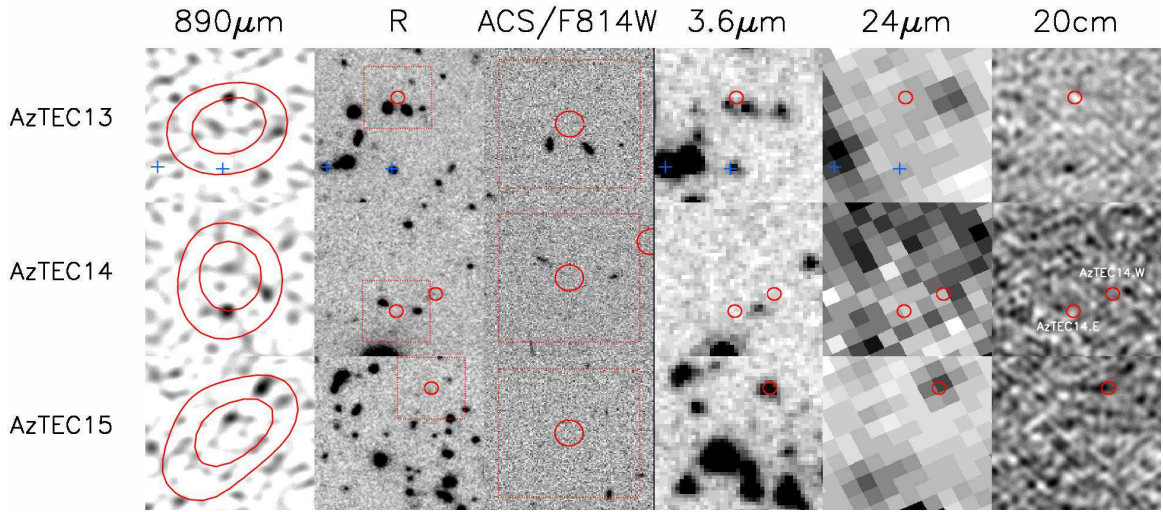}
\caption{Same as Figure~\ref{fig:stamps}.}
\label{fig:stamps2}
\end{figure*}

\begin{deluxetable*}{cccccccccc}
\tiny
\tablewidth{0pt}
\tablecaption{SMA Detections of AzTEC Sources}
\tablehead{
\colhead{} & \colhead{Name} & \colhead{$F_{\rm 1100\mu m}^{a}$} &  \colhead{$F_{890{\rm \mu m}}^{peak}$} & \colhead{S/N} & \colhead{$F_{890{\rm \mu m}}^{fit}$} &  \colhead{$\sigma(\alpha)^b$} & \colhead{$\sigma(\delta)^b$} & \colhead{AzTEC Offset} \\
\colhead{} & \colhead{} & \colhead{[mJy]} & \colhead{[mJy]} & \colhead{} & \colhead{[mJy]} & \colhead{[arcsec]} & \colhead{[arcsec]} & \colhead{[arcsec]}
}
\startdata
AzTEC8 & AzTEC J095959.34+023441.0 & $5.5^{+1.3}_{-1.3}$ & $19.7\pm 1.8$ & 10.4 & $21.6\pm 2.3$ & 0.1 & 0.1 & 4.9 \\
AzTEC9 & AzTEC J095957.25+022730.6 & $5.8^{+1.3}_{-1.5}$ & $9.0\pm 2.2$ & 4.1 & $7.4\pm 3.0$ & 0.4 & 0.4 & 1.2 \\
AzTEC10 & AzTEC J095930.76+024033.9 & $4.7^{+1.3}_{-1.3}$ & $5.3\pm 1.0$ & 5.3 & $4.7\pm 1.7$ & 0.4 & 0.4 & 0.5 \\
AzTEC11$^c$ & AzTEC J100008.91+024010.2 & $4.7^{+1.3}_{-1.3}$ & $7.4\pm 1.9$ & 8.2 & \ldots & \ldots & \ldots & \ldots \\
AzTEC11.N$^c$ & AzTEC J100008.91+024009.6 & \ldots & \ldots & \ldots & $10.0\pm 2.1$ & 0.2 & 0.2 & 2.2\\
AzTEC11.S$^c$ & AzTEC J100008.94+024012.3 & \ldots & \ldots & \ldots & $4.4\pm 2.1$  & 0.2 & 0.2 & 4.7 \\
AzTEC12 & AzTEC J100035.29+024353.4 & $4.5^{+1.3}_{-1.5}$ & $13.5\pm 1.8$ & 7.5 & $12.8\pm 2.9$ & 0.2 & 0.3 & 1.7  \\
AzTEC13 & AzTEC J095937.05+023320.0 & $4.4^{+1.3}_{-1.4}$ & $8.2\pm 1.8$ & 4.6 & $10.0\pm 2.8$ & 0.3 & 0.2 & 4.5 \\
AzTEC14.E & AzTEC J100010.03+023014.7 & $4.3^{+1.4}_{-1.4}$ & $5.0\pm 1.0$ & 5.0 & $6.1\pm 1.7$ & 0.3 & 0.3 & 5.4 \\
AzTEC14.W & AzTEC J100009.63+023018.0 & $4.3^{+1.4}_{-1.4}$ & $3.9\pm 1.0$ & 3.9 & $4.7\pm 1.7$ & 0.4 & 0.3 & 6.0 \\
AzTEC15 & AzTEC J100012.89+023435.7 & $4.2^{+1.3}_{-1.4}$ & $4.4\pm 1.0$ & 4.4 & $5.8\pm 1.7$ & 0.3 & 0.2 & 8.9 \\
\enddata
\tablenotetext{a}{Deboosted flux density at 1100\micron \citep{scott2008}.  The S/N at 1100\micron was estimated using the peak map signal, rather than the deboosted flux density.}
\tablenotetext{b}{The positional uncertainties include both the statistical and systematic uncertainties \citep[see \S~\ref{sec:overview} and][for details]{younger2007}.}
\tablenotetext{c}{AzTEC11 shows significant structure in the calibrated visibility data.  It is best modeled with two point sources, which we designate AzTEC11.N and AzTEC11.S.  In this Table, we separate out the S/N of the signal in the map and the fitted fluxes and offsets of the two components.}
\label{tab:offset}
\end{deluxetable*}

\subsection{Overview}
\label{sec:overview}

\begin{figure}
\plotone{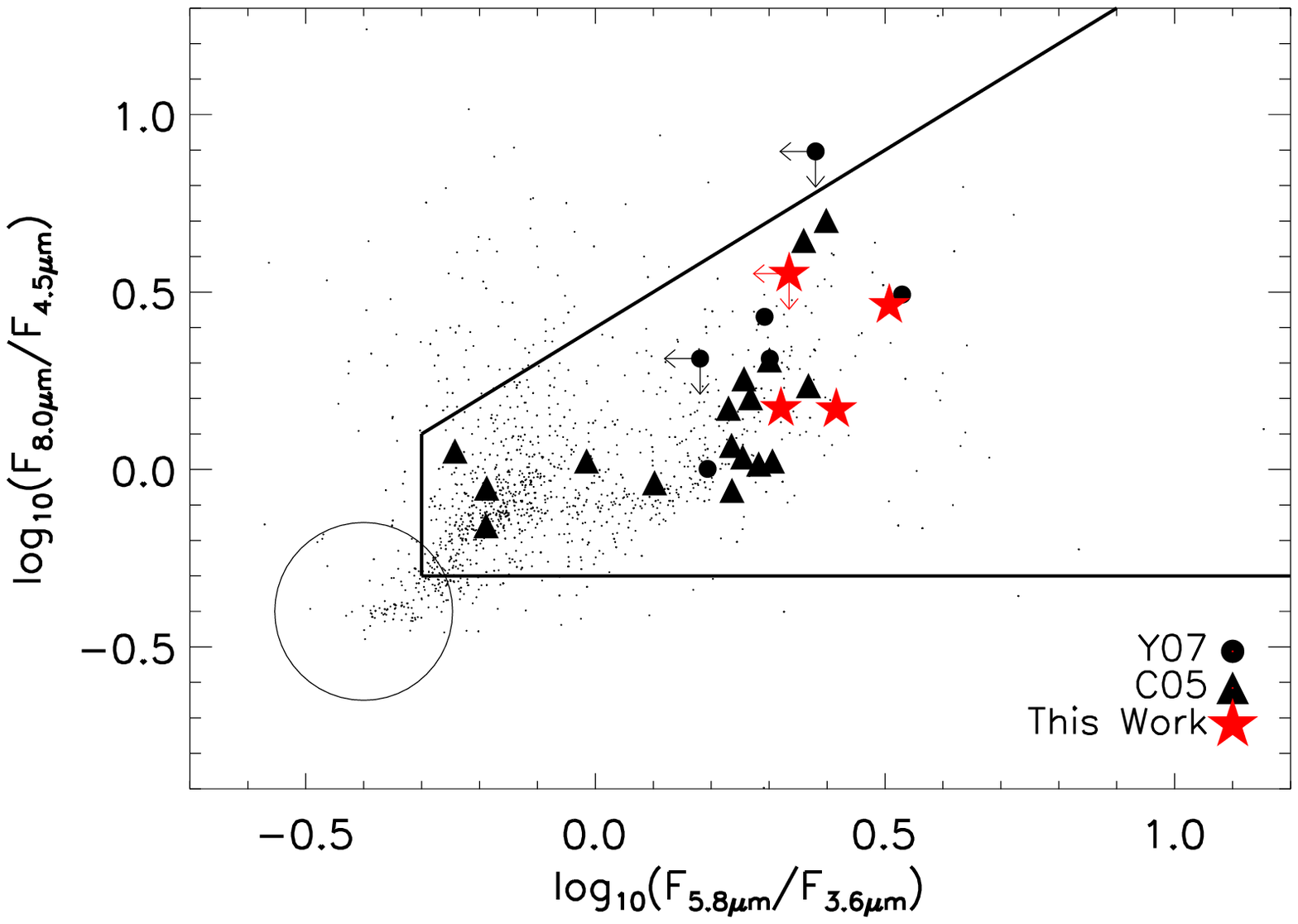}
\caption{The mid-IR colors for the sources presented in this paper (red stars; limited to those with sufficient IRAC detections) as compared to those presented in \citet[black circles:][]{younger2007}, the radio-selected sample of \citet[black triangles:][]{chapman2005}, and field sources from the HDFN.  The selection criteria proposed by \citet{yun2008} are indicated by solid lines; all of the SMGs in the sample are consistent with these color cuts.  The circle indicates the centroid of foreground sources from \citet{lacy2004}.  Arrows indicate limits.}
\label{fig:yun}
\end{figure}

The multiwavelength data -- including high-resolution 890\micron SMA imaging -- for the 8 AzTEC 1100\micronend-selected targets is summarized in Tables~\ref{tab:offset} and \ref{tab:photom}, and Figures~\ref{fig:stamps} and \ref{fig:stamps2}.  There are significant ($\gsim 3.9\sigma$) SMA detections for each of the target sources, from which we derive absolute positions accurate to $\approx 0.2-0.3$ arcsec.  Of those, all but one SMA source are detected in the IRAC 3.6 and 4.5\micron imaging (and all but two in the IRAC 5.8 and 8.0\micron imaging), while only 5 have MIPS 24\micron or radio counterparts.  We also find (see Figure~\ref{fig:yun}) that those SMA detections with sufficient IRAC coverage meet the selection criteria proposed by \citet{yun2008}.   Below, we summarize the data for each target source individually.  In what follows, we will refer to the 5/9 SMA sources detected at a peak flux density with S/N $>5$ as ``high-significance" source, while the remainder (4/9) will be referred to as ``moderate significance" and should be not considered secure detections unless they have corroborating multiwavelength counterparts; the number of beams in a typical SMA map, in addition to experience with coarsely sampled interferometer data, leads us to believe that these moderate significance detections may be spurious.  Furthermore, we make occasional reference to ``power-law" IRAC sources, which should be taken to refer to objects whose observed IRAC colors are consistent with a significant hot dust continuum thought to be powered by an active galactic nucleus \citep[AGN:][]{lacy2004,stern2005,barmby2006,donley2007,hickox2007}, ongoing intense star formation \citep{yun2008}, or a complex mix of the two \citep{younger2009.warmcold}.  

\subsection{Notes on Individual Targets}
\label{sec:individual}

{\it AzTEC J095959.34+023441.0} (AzTEC8) -- AzTEC8 is detected at high-significance (peak S/N $\approx 10$) in the SMA image.  Its visibility data is best fit by a point-source with $F_{\rm 890\mu m} =  19.7\pm 1.8$ mJy offset from the AzTEC centroid by 4.9 arcsec.  Though the offset may seem large compared to, e.g., the signal-to-noise weighted error circle \citep[$\sigma \approx 0.6\times {\rm FWHM\, (S/N)^{-1}}\sim 1.9$ arcsec for AzTEC8, see][]{ivison2007}, it is comparable to the expectation for an AzTEC source with ${\rm S/N}\geq 4.5$ \cite[4.5 arcsec at 80\% confidence, see][]{scott2008} and is still well within the 18 arcsec FWHM AzTEC beam (see also overlaid AzTEC contours in Figure~\ref{fig:stamps}). Of the two compact candidate radio counterparts (AzTEC8.E and AzTEC8.W) within the AzTEC beam, only one (AzTEC8.W; $F_{\rm 20 cm} = 89\pm11$ $\mu$Jy, and coincident with a $\sim 20$ mJy peak in the SMA map) is a significant source of submillimeter emission.  AzTEC8.E ($F_{\rm 20 cm} = 139\pm20$ $\mu$Jy) is just within the $\sim 2\sigma$ contour but not on a local maximum.  Both are power-law IRAC SEDs, and AzTEC8.E is coincident with a bright MIPS 24\micron source ($F_{\rm 24\mu m} = 820\pm 11$)\footnote{Due to the proximity of AzTEC8.W to this bright 24\micron source ($\lsim 1/2$ beam FWHM), it is difficult to obtain a reliable estimate of an upper limit on its 24\micron emission.}.  AzTEC8.W is also has faintest optical counterpart of the pair.

{\it AzTEC J095957.25+022730.6} (AzTEC9) -- AzTEC9 is detected at moderate significance (peak S/N $\approx 4$) in the SMA image.  Its visibility data is best fit by a point-source with $F_{\rm 890\mu m} =  7.4\pm 3.0$ mJy offset by 1.2 arcsec from the AzTEC centroid.  This fitted flux density, while lower than that inferred from the image plane, is consistent with its peak flux assuming a point-source structure.  It is coincident with IRAC 3.6 and 4.5\micronend, and a radio source with $F_{\rm 20cm}=59\pm 10$ $\mu$Jy, but it is not detected in optical, IRAC 5.8/8.0\micronend, or MIPS 24\micron imaging data.  Therefore, despite the moderate significance of the detection, because it is coincident with an IRAC/radio source we believe the detection is real. 

\begin{deluxetable*}{ccccccccccccc}
\tablewidth{0pt}
\tablecaption{Multi-Wavelength Photometry}
\tablehead{
\colhead{Name} & \colhead{B$^a$} & \colhead{r$^a$} & \colhead{$F_{\rm 3.6\mu m}^b$} & \colhead{$F_{\rm 4.5\mu m}^b$} & \colhead{$F_{\rm 5.8\mu m}^b$} & \colhead{$F_{\rm 8.0\mu m}^b$} & \colhead{$F_{\rm 24\mu m}^c$} & \colhead{$F_{\rm 20cm}^d$} \\
 & [mag] & [mag] & [$\mu$Jy] & [$\mu$Jy] & [$\mu$Jy]   & [$\mu$Jy] & [$\mu$Jy] & [$\mu$Jy] & 
}
\startdata
AzTEC8.E$^e$ & $25.69\pm 0.11$ & $25.00\pm 0.07$ & $10.8\pm 0.1$ & $14.8\pm 0.2$ & $28.3\pm 0.9$ & $69.6\pm 2.5$ & $820\pm 11$ & $139\pm 20$ \\
AzTEC8.W$^e$ & $>27.8$ & $>27.2$ & $7.4\pm 0.1$ & $11.9\pm 0.2$ & $23.8\pm 0.9$ & $34.6\pm 2.6$ & \ldots & $89\pm 11$ \\
AzTEC9 & $>27.8$ & $>27.2$  & $2.5\pm 0.1$ & $3.2\pm 0.2$ & $< 11.3$ & $<14.6$ & $<71$ & $59\pm 10$ \\
AzTEC10 & $>27.8$ & $>27.2$  & $7.1\pm 0.1$ & $11.7\pm 0.2$ & $18.5\pm 0.9$ & $17.3\pm 2.3$ & $114\pm 16$ & $<33$ \\
AzTEC11$^f$ & $>27.8$ & $>27.2$   & $29.9\pm 0.2$ & $42.3\pm 0.3$ & $50.7\pm 1.1$ & $42.1\pm 2.5$ & $644\pm 11$ & \ldots \\
AzTEC11.N$^f$ & $23.91\pm 0.04$ & $23.95\pm 0.04$ & \ldots & \ldots & \ldots & \ldots & \ldots & $120\pm 26$ \\
AzTEC11.S$^f$ & $>27.8$ & $>27.2$  & \ldots & \ldots & \ldots & \ldots & \ldots & $115\pm 26$ \\
AzTEC12 & $25.73\pm 0.12$ & $24.83\pm 0.06$ & $28.5\pm 0.2$ & $38.2\pm 0.3$ & $59.6\pm 1.1$ & $56.9\pm 2.4$ & $344\pm 11$ & $104\pm 14$ \\
AzTEC13 &  $>27.8$ & $>27.2$  & $<0.9$ & $<1.7$ & $<11.3$ & $<14.6$ & $<71$ & $<33$ \\
AzTEC14.E & $>27.8$ & $>27.2$  & $<0.9$ & $<1.7$ & $<11.3$ & $<14.6$ & $<71$ & $<33$ \\
AzTEC14.W & $>27.8$ & $>27.2$  & $<0.9$ & $<1.7$ & $<11.3$ & $<14.6$ & $<71$ & $<33$ \\
AzTEC15  & $>27.8$ & $>27.2$  & $7.2\pm 0.1$ & $12.1\pm 0.2$ & $13.8\pm 0.9$ & $27.1\pm 2.5$ & $76\pm 11$ & $<33$ \\
\enddata
\tablenotetext{a}{B and r' optical imaging data is measured in a 2 arcsec aperture.  Upper limits are at the 3-$\sigma$ level \citep{taniguchi2007}.}
\tablenotetext{b}{IRAC 3.6-8\micron fluxes are measured in a 2 arcsec aperture with the appropriate aperture correction.  Upper limits are at the 5-$\sigma$ level \citep{sanders2007}.}
\tablenotetext{b}{MIPS 24\micron fluxes are measures in a $7\times 7$ arcsec square aperture, with the appropriate aperture correction.  Upper limits are at the 5-$\sigma$ level \citep{sanders2007}.}
\tablenotetext{d}{VLA fluxes are a Gaussian fit to the imaging data.  They do not include corrections for bandwidth smearing, which will raise them approximately 15\%.  Upper limits are at the 3-$\sigma$ level.}
\tablenotetext{e}{This source has two candidate radio counterparts with comparable flux density at 20cm: AzTEC8.E and AzTEC8.W.  However, only one was detected by the SMA.  Of the two candidates, one is a bright MIPS 24\micron source and is not associated with the submm emission (AzTEC8.E), while the other is not (AzTEC8.W).  This source is similar to LH850.02 from \citep{younger2008}.}
\tablenotetext{e}{AzTEC11 shows significant structure in the calibrated visibilities that are best modeled with a double point source.  Optical imaging data also suggests a two-component structure, with the northern source (AzTEC11.N) having an optical counterpart with $z_{phot} = 2.55$.  IRAC and MIPS flux measurements are likely a blend of both sources.  A double-gaussian model has been fit to the VLA imaging data, yielding individual fluxes for each component.}
\label{tab:photom}
\end{deluxetable*}

{\it AzTEC J095930.76+024033.9} (AzTEC10) -- AzTEC10 is detected at moderate significance (peak S/N $\approx 5$) in the SMA image.  Its visibility data is best fit by a point-source with $F_{\rm 890\mu m} =  5.3\pm 1.0$ mJy offset by 0.5 arcsec from the AzTEC centroid.  The SMA detection is coincident with an IRAC 3.6-8.0\micron source which peaks at 5.8\micronend -- though it is statistically consistent with a power-law SED -- and a faint MIPS source with $F_{\rm 24\mu m} = 114\pm 16$ $\mu$Jy.  AzTEC10 is not detected at optical or radio wavelengths.   

{\it AzTEC J100008.91+024010.2} (AzTEC11) -- AzTEC11 is detected at high-significance (peak S/N $\approx 8$) in the SMA image.  Its visibility data best fit by a double point-source, the two components of which (AzTEC11.N and AzTEC11.S) have $F_{\rm 890\mu m}=10.0\pm 2.1$ and $4.4\pm 2.1$ mJy and are offset by 2.2 and 4.7 arcsec from the AzTEC centroid respectively.  There is an extended IRAC and MIPS source in between the two detections which is likely a blend of both sources.  In addition, there is an elongated radio source that is extended in the same direction as the submillimeter detection; a double gaussian fit to the radio image data yields two sources with integrated flux densities of $F_{\rm 20cm} = 120\pm 26$ and $115\pm 26$ $\mu$Jy coincident with AzTEC11.N and AzTEC11.S respectively.  Furthermore, AzTEC11.N has a photometric redshift of $z_{phot}=1.78^{+0.15}_{-0.23}$ and a fitted stellar mass of ${\rm log}(M_\star/M_\odot)=10.9$ \citep{mobasher2006}.

{\it AzTEC J100035.29+024353.4} (AzTEC12) -- AzTEC12 is detected at high-significance (peak S/N $\approx 8$) in the SMA image.  The visibility data is best fit by a point-source with $F_{\rm 890\mu m} =  12.8\pm 2.9$ mJy offset by 1.7 arcsec from the AzTEC centroid.   As with AzTEC10, the SMA detection is coincident with an IRAC 3.6-8.0\micron source which peaks at 5.8\micron, indicative of $2\lsim z \lsim 3.5$ for a starburst \citep{huang2004}, in addition to a bright MIPS 24\micron source ($F_{\rm 24\mu m}=344\pm 11$ $\mu$Jy) and a radio counterpart with $F_{\rm 20cm} = 104\pm 14$ $\mu$Jy.

{\it AzTEC J095937.05+023320.0} (AzTEC13) -- AzTEC13 is detected at moderate significance (peak S/N $\approx 5$) in the SMA image.  The visibility data is best fit by a point-source with $F_{\rm 890\mu m} =  10.0\pm 2.8$ mJy offset by 4.5 arcsec from the AzTEC centroid.   It is not coincident with any optical, IRAC, MIPS, or radio sources.  There is also a radio source within half an AzTEC beam (6.7 arcsec from the AzTEC centroid) with an IRAC 3.6\micron detection and a photometric redshift of $z_{phot}=0.72^{+0.02}_{-0.07}$ \citep{mobasher2006} which is not associated with the submillimeter emission -- the 890\micron emission at its location is consistent with the noise level of the SMA map.  

{\it AzTEC J100010.03+023014.7} (AzTEC14.E) -- AzTEC14.E is a $\approx 5\sigma$ peak in the SMA image.  Its visibility data is best fit by a point-source with $F_{\rm 890\mu m} =  6.1\pm 1.7$ mJy offset by 5.4 arcsec from the AzTEC centroid.  It is not coincident with any optical, IRAC, MIPS, or radio sources.

{\it AzTEC J100009.63+023018.0} (AzTEC14.W) -- AzTEC14.E is a $\approx 4\sigma$ peak in the SMA image.  Its visibility data is best fit by a point-source with $F_{\rm 890\mu m} =  4.7\pm 1.7$ mJy offset by 6.0 arcsec from the AzTEC centroid.  It is not coincident with any optical, IRAC, MIPS, or radio sources.

{\it AzTEC J100012.89+023435.7} (AzTEC15) -- AzTEC15 is detected at moderate significance (peak S/N $\approx 4$) in the SMA image.  Its visibility data is best fit by a point-source with $F_{\rm 890\mu m} =  5.8\pm 1.7$ mJy offset by 8.9 arcsec from the AzTEC centroid.  As with AzTEC8, this is well beyond the formal error radius estimated by both \citet{scott2008} and \citet{ivison2007}, and is in fact at the edge of the AzTEC beam FWHM.  It is, however, coincident with a power-law IRAC sources and a faint MIPS 24\micron source ($F_{\rm 24\mu m} = 76\pm 11$ $\mu$Jy), but is not detected at optical or radio wavelengths.  Though it only $\approx 4\sigma$ and offset by $\sim 1/2$ an AzTEC beam from the 1100\micron centroid, we believe that detections by IRAC and MIPS indicate that the source is real.

\section{Discussion}
\label{sec:discuss}

The data presented in this work, combined with those presented by \citet{younger2007}, constitute an unbiased sample of 15 AzTEC 1100\micronend-selected sources with complete submillimeter interferometric followup.  Positions derived from the interferometric imaging provide positions accurate to $\lsim 0.3$ arcsec at a resolution (beam size $\approx 2$ arcsec FHWM) that is well-matched to multiwavelength imaging data.  This, combined with the rich multiwavelength dataset available for the COSMOS field \citep[see \S~\ref{sec:obs} and][for an overview]{scoville2007} immediately enables three investigations: in \S~\ref{sec:radio.submm} we present a systematic test of the radio-submillimeter association, in \S~\ref{sec:morphology} we examine the rest frame far-IR morphology of the sample, and in \S~\ref{sec:highz} we identify candidate high-redshift $z\gsim 3$ sources.   In the following analyses, we utilize the combined results of this work and \citet{younger2007} for a total sample of 15 1.1mm selected SMGs with complete interferometric followup.  As noted in \S~\ref{sec:overview}, sources at the S/N $\approx 4-5$ level with no confirming detection at other wavelengths may be spurious.  This does not, however, effect our conclusions.

\subsection{Testing the Radio-Submillimeter Association}
\label{sec:radio.submm}

\begin{figure}
\plotone{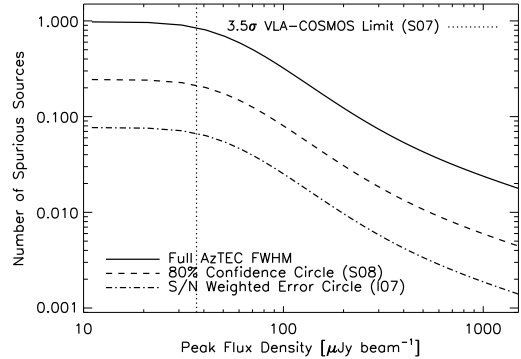}
\caption{The expected number of spurious sources associated with AzTEC positions for all 15 sources in the sample.  We consider three different definitions of the search area around AzTEC detections: (solid line) the full AzTEC beam FWHM of 18 arcsec, (dashed line) the 80\% confidence separation of 4.5 arcsec as estimated by \citet[S08:][]{scott2008}, and (dot-dashed line) the expected positional uncertainty of fitted point-sources to a radio map $\sigma \approx 0.6\times {\rm FWHM\, (S/N)^{-1}}$ \citep[I07:][]{ivison2007}.  The number counts were taken from the polynomial fitting function of \citet{bondi2008}, and agree with results from other deep 1.4 GHz radio surveys \citep[e.g.,][]{hopkins2003}.  The dotted line indicates the $3.5\sigma$ limiting flux density of the VLA-COSMOS survey \citep[S07:]{schinnerer2007}.  Even under the most conservative assumption of the full AzTEC beam, we would expect at most $\lsim 1$ spurious radio source in the full sample of 15 objects.}
\label{fig:spurious}
\end{figure}

The radio-submillimeter association uses 20cm radio sources imaged at comparatively higher resolution \citep[20cm beam size $\approx 1$ arcsec, absolute astrometric uncertainty $\approx 0.1$ arcsec: e.g.,][]{ivison2002} to refine absolute position measurements for SMGs.   This technique leverages the local far-IR radio relation, combined with statistical arguments, to associate SMGs with radio sources within the submillimeter beam \citep[see also][]{pope2006,ivison2007} .  While this method is efficient and physically plausible given the apparent lack of significant evolution in the far-IR/radio correlation with redshift \citep[e.g.,][]{ibar2008,younger2008.egsulirgs}, an independent verification of this technique is important.  Our unbiased sample of 15 SMGs with direct submillimeter positions from interferometric imaging is the first robust sample for such a study.  

From a statistical standpoint, we expect the contamination from spurious radio detections to be minimal.  Though some observations reveal excess clustering of sub-mJy radio sources on arcminute scales \citep[e.g.,][]{richards2000,georgakakis2000}, those same observations show no evidence of significant anisotropy in 50-100 $\mu$Jy radio sources on arcsecond scales of the kind considered here.  Therefore, we can estimate the expected number of spurious associations with AzTEC positions assuming a roughly uniform distribution of foreground radio sources and the polynomial fitting form for radio number counts in the VLA-COSMOS survey as measured by \citet{bondi2008}.  The results of this exercise are summarized in Figure~\ref{fig:spurious}.  We consider three definitions of the search radius around AzTEC positions: (1) the full 18 arcsec AzTEC beam FWHM at 1.1mm, (2) the 4.5 arcsec 80\% confidence interval for high-significance (${\rm S/N} > 4.5$) AzTEC detections as estimated via Monte-Carlo simulation \citep{scott2008}, and (3) the signal-to-noise weighted error circle of $\sigma \approx 0.6\times {\rm FWHM\, (S/N)^{-1}}$ \citep{condon1997,ivison2007}.  Owing to their relatively low surface density, even under the most conservative assumption of a full 18 arcsec search radius, we would expect $\lsim 1$ spurious association between AzTEC and radio detections in the entire sample of 15 objects, and under more realistic assumptions none at all.

\begin{figure*}
\plottwo{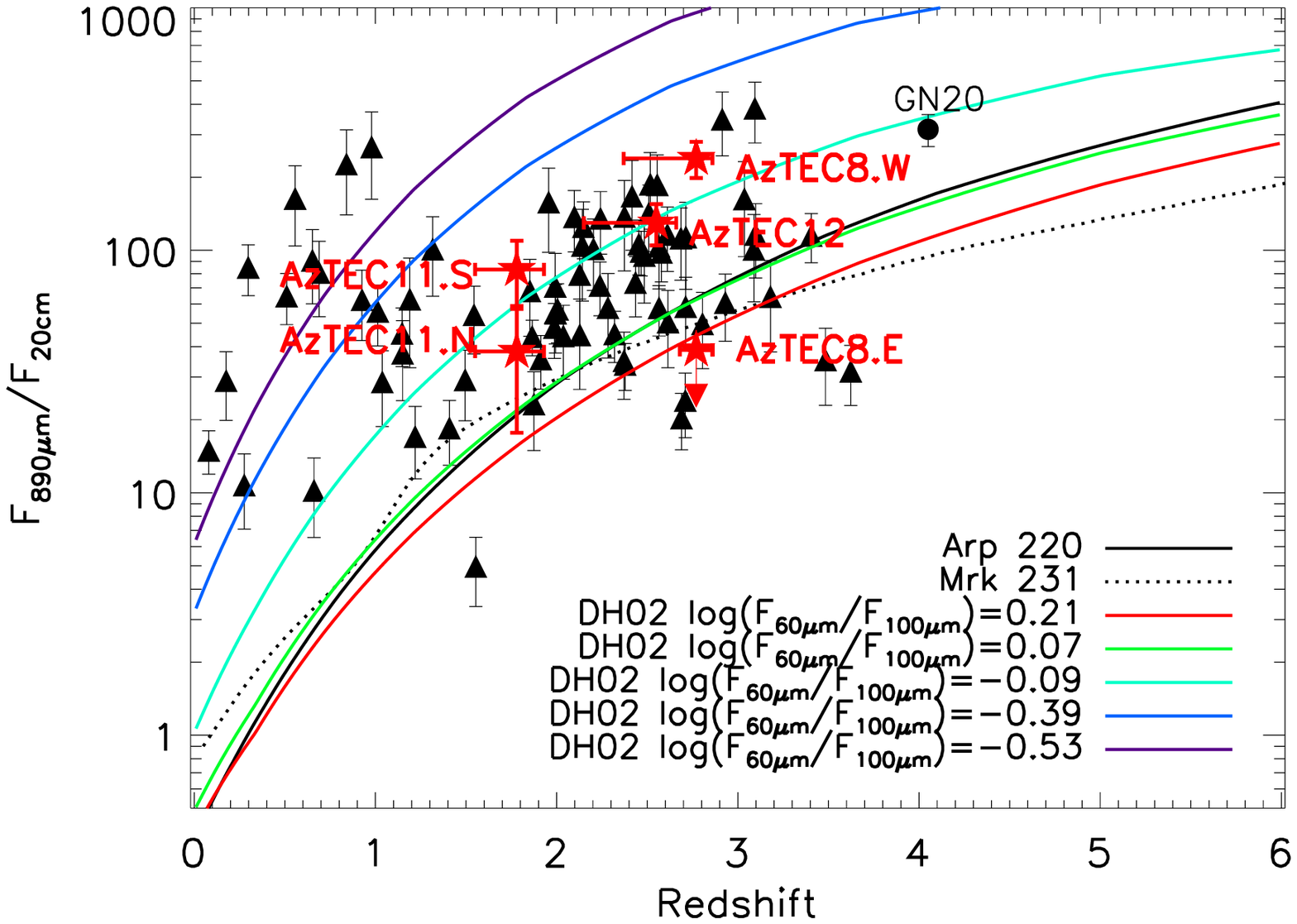}{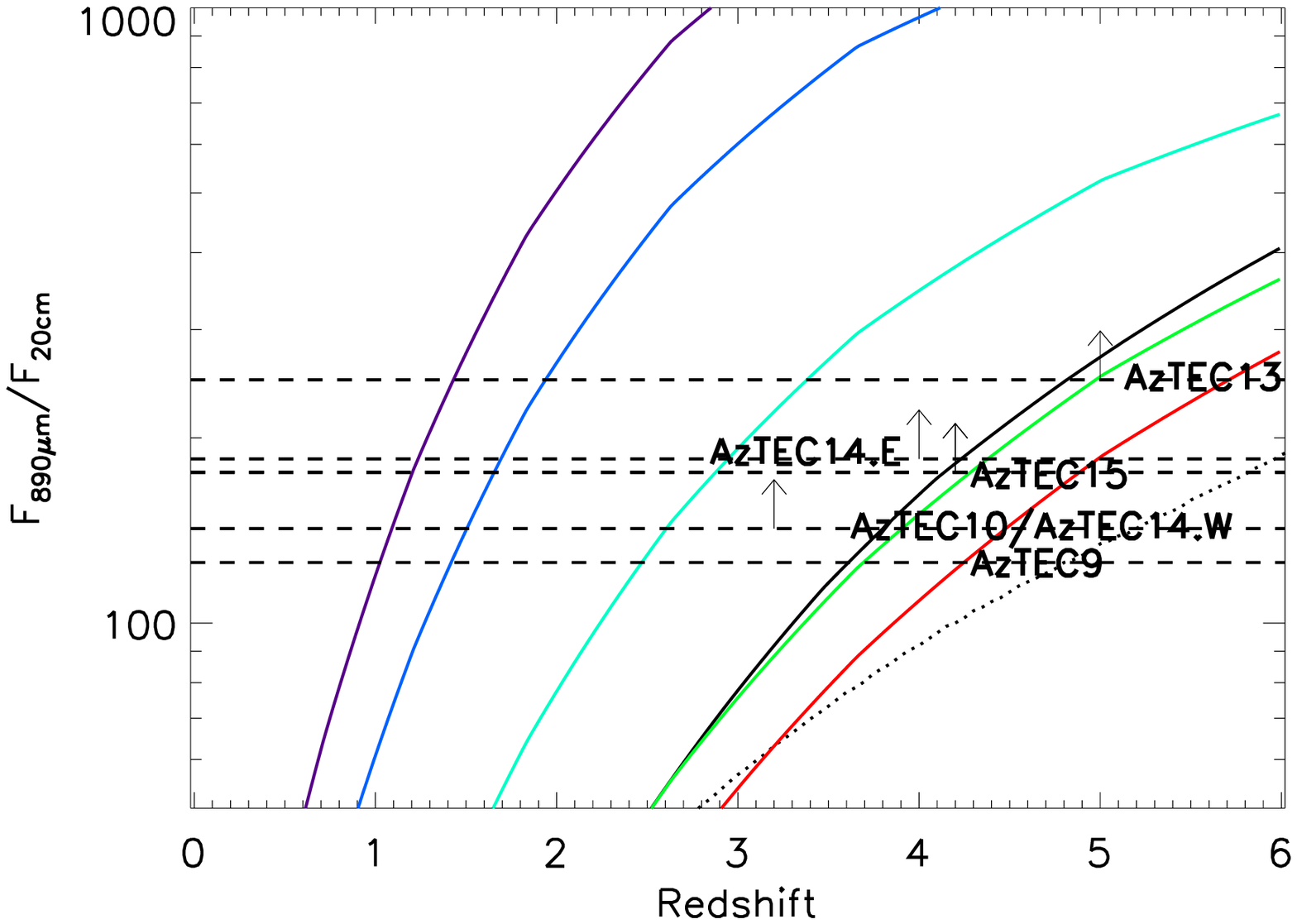}
\caption{The observed submillimeter-to-radio flux density ratio for the sample and radio-selected SMGs \citep[solid triangles:][]{chapman2005}.  The location of GN20, a spectroscopically confirmed $z\sim 4$ SMG \citep{daddi2008} is shown as a solid circle.  The solid colored lines show the range of model tracks from \citet{dale2002}, the solid line is Arp 220 (a typical local ULIRG), and the dotted line is Mrk 231, which shares many characteristics with Seyfert 1 sources and radio-quiet quasars \citep[e.g.,][]{maloney2000}.  The left panel shows sources in the sample with photometric redshifts (red stars), while the right panel is a zoom-in of the remaining sources (dashed lines), and arrows indicate limits.  We find that a number of objects in the sample -- AzTEC 9, 10, 13, 14.W, 14.S, and 15 -- have submillimeter-to-radio flux density ratios consistent with a higher median/average redshift than radio-selected samples.}
\label{fig:radio}
\end{figure*}

A complementary method of estimating contamination by foreground radio sources is the $P$-statistic \citep{downes1986,dunlop1989,scott1989}.  For a given cumulative distribution of number counts as a function of flux density $n_c(>S)$, one can estimate the probability of a random association for a source with flux density $S$ and separation $\theta$ as $P = 1-{\rm exp}(-\pi n_c \theta^2)$.  This method has been used by a number of authors to argue for the likely association between SMGs and radio or mid-IR sources proximate to the submillimeter position \citep[e.g.,][]{hughes1998,lilly1999,ivison2002,pope2006,ivison2007}.  For all the sources in the sample, and again assuming the number counts of \citet{bondi2008}, owing to their low surface density on the sky the random association of any significant radio source (S/N $\gsim 3.5$) within an individual AzTEC beam FWHM of 18 arcsec is highly improbable $P < 5\%$.  Therefore, we would expect $\lsim 1$ total spurious association for the entire sample, and far less under more realistic assumptions for AzTEC positional uncertainty (see above).  Furthermore, considering the specific radio detections and their separation from the AzTEC centroid, the likelihood of a random superposition is even lower, with $P << 1\%$.  

We find that of the 9 sources in the sample with $\gsim 3\sigma$ radio sources within the AzTEC beam, there is only one instance -- AzTEC13 -- in which none of the radio detections within the AzTEC beam is also detected in the high-resolution SMA submillimeter maps; a result consistent with the statistics estimated above.  This target has a $8.2\pm 1.8$ mJy peak 4.5 arcsec to the north of the AzTEC centroid, but no significant detection at the location of the radio source 6.7 arcsec to the south.  This radio source has a 20cm flux density of $66\pm 10$ $\mu$Jy, a photometric redshift of $z_{phot} = 0.72^{+0.02}_{-0.07}$, and is detected at 3.6\micron by IRAC.  It furthermore has a very high probability of being associated with the AzTEC emission; considering its peak flux density and separation from the AzTEC centroid we estimate $P\approx 2\%$. By contrast, the submillimeter source is not detected at any other wavelength.  This is quite surprising, given the frequency of bright SMGs with unambiguous position measurements and 3.6\micron counterparts \citep{iono2006,younger2007,younger2008,wang2007,wang2008}.  Without additional information, including deeper radio and IRAC imaging, we can only speculate that this source either lies at extremely high-redshift and/or has an uncharacteristically low stellar mass $\lsim 10^{11}M_\odot$ \citep[][]{borys2005}.  At the same time, though we believe this detection is reliable, we must also admit the possibility -- for the reasons discussed in \S~\ref{sec:discuss} -- that an SMA detection at S/N $\approx 5$ without confirming detections at other wavelengths is spurious.

\subsection{The Rest-Frame Far-IR Morphology of SMGs}
\label{sec:morphology}

In addition to providing accurate absolute position measurements, the interferometric imaging constrains the submillimeter morphology of bright SMGs.   This allows us to both identify multi-component sources and estimate the physical scale of the far-IR.  The former constrains the contribution of blends to the bright SMG population, and the latter is related to the engine driving the tremendous luminosity of these sources \citep[see also the discussion in ][]{younger2008highres}.

Our sample of 15 targets contains two reliable targets\footnote{As noted in \S~\ref{sec:discuss}, sources S/N $\lsim 4-5\sigma$ without confirming detections at other wavelengths are possibly spurious.  For this reason, we exclude AzTEC14.W. and E from this discussion, though if the source positions are confirmed with more sensitive measurements this source may represent an example of a truly blended source.} with multiple radio detections within the AzTEC beam: AzTEC5 (AzTEC5.N and AzTEC5.S) and AzTEC8 (AzTEC8.E and AzTEC8.W).   In both cases, SMA imaging identifies only one of the two radio components as the source of the submillimeter emission.  In both instances, the radio pairs have photometric redshifts that are marginally consistent with the sources residing at the same redshift, and therefore may be physically associated; AzTEC5.W \citep[the submillimeter source; see Figure~1 in][]{younger2007} has $z_{phot}=1.50^{+0.19}_{-0.10}$ and AzTEC5.E ($\approx 7$ arcsec away) has $z_{phot}=2.95^{+0.05}_{-1.13}$ (more appropriate, these photometric redshifts are not inconsistent to within the stated statistical errors); AzTEC8.W (the submillimeter source; see Figure~\ref{fig:stamps}) has a submillimeter-to-radio flux density ratio consistent with $z\approx 2-3$ (see Figure~\ref{fig:radio}) and AzTEC8.E ($\approx 2$ arcsec away) has $z_{phot}=2.77^{+0.09}_{-0.40}$.  This is consistent with the statistics of radio sources in the VLA-COSMOS survey; under the same assumptions as \S~\ref{sec:radio.submm}, the probability of the secondary radio counterparts in AzTEC5 and AzTEC8 constituting a random superposition of foreground sources are both $<<1\%$.  Therefore, both objects are similar to LH850.02, a bright SMG in the Lockman Hole with two candidate radio counterparts of which only one was associated with the submillimeter emission \citep{younger2008} and support the predicted rarity of SMGs arising from confusion of two lower luminosity sources \citep{ivison2007}.  If these two objects are indeed at the same redshift, then they necessarily have significantly different far-IR SEDs -- or equivalently, effective dust temperatures -- with the submillimeter-detected source representing a relatively cold starburst \citep[$T_{dust} \approx 30-40$ {\sc K};][]{kovacs2006,coppin2008.sharcii,younger2008.egsulirgs}, and the other a warm starburst or IR-luminous AGN \citep{sanders1996,lutz1998,risaliti2000,younger2009.warmcold,casey2009}.

Of the remaining sources, all but one are best modeled as unresolved point-sources, which constrains their apparent angular size to $\lsim 1.2$ arcsec.  Furthermore, the one source that is resolved by the SMA -- AzTEC11 -- is best modeled as a double point-source, which is consistent with the two components of the extended radio emission within the AzTEC beam (see \S~\ref{sec:individual}, Table~\ref{tab:photom} and Figure~\ref{fig:stamps}), and its visibility function is inconsistent with extended emission such as a Gaussian or disk morphology at the $\sim 2-3\sigma$ level.  Therefore, all the SMA detections in the unbiased sample of 15 targets are compact in the SMA imaging data, which at $z\sim 2-3$ corresponds to a physical scale of $\lsim 9-8$ kpc.  While only an upper limit, these sizes are consistent with those measured via higher resolution submillimeter \citep{younger2008highres} and radio \citep{chapman2004,biggs2008} data, and rule out cool cirrus dust \citep{efstathiou2003,kaviani2003} on $>10$ kpc scales as the source of the far-IR luminosity in the majority of bright SMGs.  However, higher resolution imaging will be required to measure the size of the starburst region, and the data cannot rule out a significant contribution from an IR-luminous AGN on sub-kpc scales.

\subsection{Candidate High-Redshift Sources}
\label{sec:highz}

\begin{figure}
\plotone{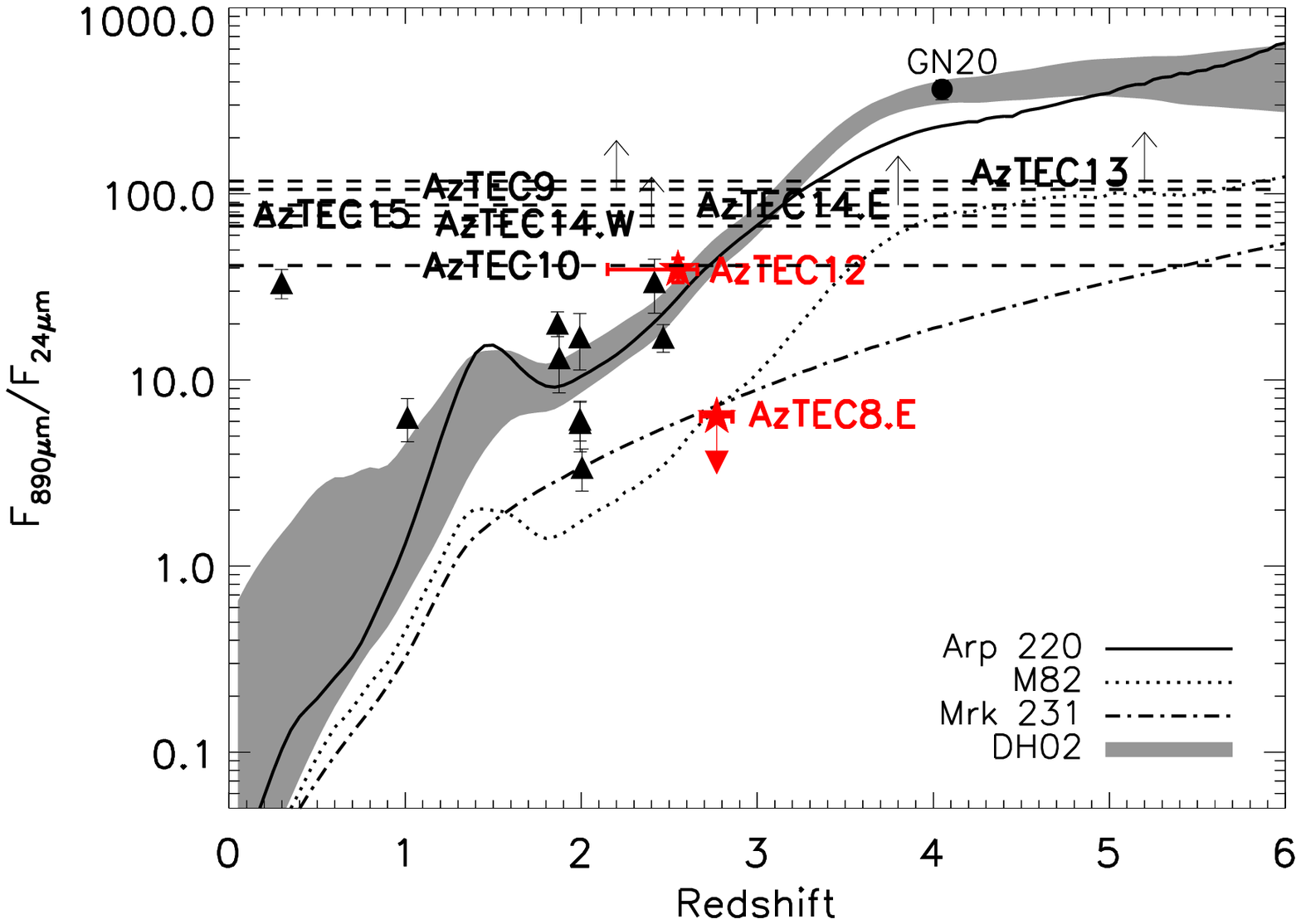}
\caption{The observed submillimeter-to-24\micron flux density ratio for the sample and radio-selected SMGs \citep[solid triangles:][]{chapman2005}.  The location of GN20, a spectroscopically confirmed $z\sim 4$ SMG \citep{daddi2008} is shown as a solid circle.  The grey shaded region shows the range of model tracks from \citet{dale2002}, the solid line is Arp 220 (a typical local ULIRG), the dotted line is M82 (a local starburst with hotter dust), and the dot-dashed line is Mrk 231 (a ULIRG with a significant AGN contribution).  Red stars indicate sources in the sample with photometric redshifts (AzTEC8.W and AzTEC12), while the rest are shown as dashed black lines, and arrows indicate limits.  As with Figure~\ref{fig:radio}, a number of objects in the sample have submillimeter-to-24\micron flux density ratios indicative of high-redshift ($z\gsim 3$).}
\label{fig:mips}
\end{figure}

\begin{figure}
\plotone{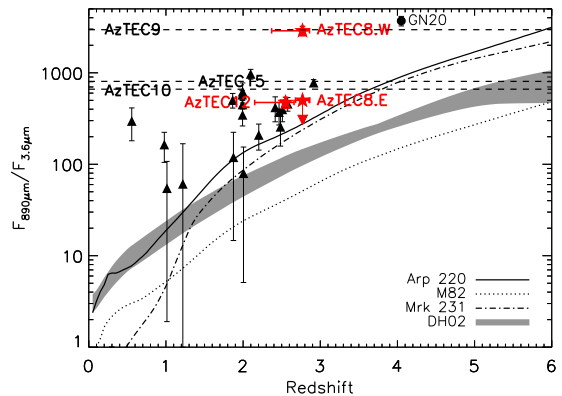}
\caption{Same as Figure~\ref{fig:mips}, but for the observed submillimeter-to-3.6\micron flux density ratio and limited to those with 3.6\micron detections.  Of those with IRAC detections, AzTEC9, 10, and 15 have flux density ratios indicative of high-redshift ($z\gsim 3$).  However, the model disagreement indicates that SMGs have significantly higher dust obscuration than is included the model calculations.}
\label{fig:irac}
\end{figure}

Though a number of techniques have been proposed \citep{ivison2002,pope2006,ashby2006,yun2008}, the unambiguous identification of multiwavelength counterparts to SMGs represents one of the most important challenges to their detailed study.  Owing to the strong $k$-correction at radio wavelengths, this problem is particularly acute for high-redshift $z\gsim 3$ SMGs.   Recently, \citet{younger2007} found evidence for a population -- five of seven AzTEC targets -- of $z\gsim 3$ SMGs with very faint or no radio or 24\micron counterparts.  The existence of a large number of these high-z SMGs has profound consequences for models of galaxy formation \citep{baugh2005,swinbank2008,coppin2009} -- which struggle to reproduce these sources, even with the arguably extreme assumption of a flat initial mass function (IMF) in starbursts -- and dust production models \citep[e.g.,][]{gehrz1989,marchenko2006,dunne2003} given the limited coordinate time since the Big Bang at these redshifts \citep[see also the discussion in \S~5 of][]{younger2007}.  Though the high-redshift of four individual sources has been confirmed spectroscopically \citep{capak2008,daddi2008,daddi2009b,schinnerer2008,coppin2009}, the role of these objects among bright SMGs, and their contribution to the comoving star formation rate density (SFRD) in the early universe, has yet to be determined.  Our expanded, unbiased sample of SMGs with interferometric followup can provide more powerful constraints, and identifies five additional candidate high-resdshift SMGs.

In Figure~\ref{fig:radio}, we present the submillimeter-to-radio flux density ratio for the new objects the sample, along with SMGs with spectroscopic redshifts and radio detections from \citet{chapman2005}.  The combined effects of a negative $k$-correction in the submillimeter \citep{blain1993} and a strong positive $k$-correction in the radio make this quantity a strong function of redshift \citep{carilli1999,yun2002}.  Those objects in the sample with photometric redshifts -- AzTEC8, 11, and 12 -- are all consistent with results from radio-selected samples.  However, those with no radio detection -- AzTEC10, 13, 14.E, 14.W, and 15 -- have submillimeter-to-radio ratios that suggest either a higher average/median redshift than radio selected samples ($z\gsim 3$) or a colder dust temperature.  However, dust temperatures cold enough to yield high flux density ratios ($F_{\rm 890\mu m}/F_{\rm 20cm} \gsim 150$) could be expected to have extended submillimeter morphologies on scales of several arcsec \citep{efstathiou2003,kaviani2003}, which is inconsistent with the observed sizes of the objects in the sample (see previous section).  Therefore, as with the radio-dim objects in \citet{younger2007}, we believe these objects are compelling candidate members of a population of high-redshift sources at $z\gsim 3$.

The mid- and near- IR properties of these objects are also consistent with a high-redshift.  In Figure~\ref{fig:mips} we present the submillimeter-to-24\micron flux density ratio, which is also a strong function of redshift \citep[see also][]{wang2007,wang2008} -- though it is important to note that the detectability of SMGs at 24\micron depends on the relative strength of PAH emission- and silicate absorption-features that are redshifted into the MIPS 24\micron band in sources at $z \sim 2$, which are quite model-dependent.   Three of the four high-redshift candidates are not detected in deep 24\micron imaging, which is consistent with $z\gsim 3$.  AzTEC10, however, has a $\approx 100$ $\mu$Jy 24\micron counterpart which, if we assume an Arp 220 template, suggests a redshift of $\sim 2-3$; more in line with the radio-selected population.  However, this object has a 3.6-8.0\micron SED consistent with a power-law, which suggests either a significant hot dust contribution from either an AGN (i.e., Mrk 231) or starburst \citep[i.e., M82; see also][]{yun2008}, which is consistent with higher redshift $z\gsim 3$.  The observed submillimeter-to-3.6\micron flux density ratio (see Figure~\ref{fig:irac}), which is also a strong function of redshift given a fairly narrow range in stellar mass and extinction \citep[e.g.,][]{borys2005}, for these five high-redshift candidates is consistent with $z\gsim 3$.   A full analysis of the near- through far-IR SED, and its implications for the starburst properties and redshift of these sources will appear in forthcoming work (M. Yun et al., in preparation).

Finally, we can use the observed frequency of high-redshift SMGs in the sample -- ten total candidate sources -- to estimate a lower limit on their contribution to cosmic star formation at $z\gsim 3$.  As a crude approximation, we assume that the ten high-redshift candidates in the sample are uniformly distributed from $3<z<5$ and have a far-IR SED similar to Arp 220, which yields a bolometric correction of $L_{FIR} = 2\times (F_{\rm 890\mu m}/{\rm mJy})\,\, 10^{12}\,\, L_\odot$ \citep{neri2003,younger2008highres}.  Given the assumed cosmology and a \citet{salpeter1955} IMF, this yields an SMG contribution of ${\rm SFRD_{SMG}}(3<z<5)\gsim 5\times 10^{-3}$ $M_\odot\, {\rm yr^{-1} Mpc^{-3}}$.  This is $\lsim 10\%$ of estimates of the universal SFRD at similar redshifts \citep{madau1996,barger2000,hopkins2004}.  That this is somewhat higher than previous estimates \citep{barger2000,ivison2002} is not surprising; the sample presented in this work contains much brighter objects than typical SCUBA 850\micron surveys (median 850\micron flux density $\approx 9-10$ mJy versus $\approx 6$ mJy for the Chapman et al. [2005] sample), and some have speculated \citep[e.g.,][]{dunlop2001,ivison2002,younger2007} that the brightest SMGs lie preferentially at higher redshift.  Therefore, we find that while bright high-z SMGs are important in constraining models of galaxy formation, the highest luminosity millimeter sources ($F_{\rm 1100\mu m} \gsim 4$ mJy; $L_{FIR}>>10^{13}$ $L_\odot$) do not dominate star formation at $3\lsim z \lsim 5$.  

\section{Conclusion}
\label{sec:conclusion}

We present results from an extensive campaign to follow up millimeter-selected SMGs with high-resolution (beam size $\sim 2$ arcsec) interferometric imaging.  In this work, we have targeted 8 high-significance AzTEC 1.1mm sources with the SMA at 890\micronend, resulting in six reliable  -- S/N $>5$ (``high-significance") or S/N $>4$ (``moderate significance") with multiwavelength counterparts -- and two tentative -- moderate significance with no multiwavelength counterparts -- detections.  From the high-resolution maps, we derived positions accurate to $\sim 0.2$ arcsec, in addition to 890\micron fluxes and size constraints. When combined with results from \citet{younger2007} this constitutes an unbiased sample of millimeter-selected SMGs with complete interferometric followup.  From this combined sample, we firstly empirically verify the radio-submillimeter association invoked by previous authors.  Second, we find that when there are two candidate radio counterparts, high-resolution imaging tends to single out one of the radio sources as the origin of the submillimeter emission, though there is some evidence that the two objects are physically associated \citep[see also][]{younger2008}.  Third, with the exception of one source (AzTEC11), all of the SMGs in the sample are unresolved by the SMA in compact configuration, yielding a maximum angular size of $\theta\lsim 1.2$ arcsec which, at redshifts typical of SMGs, is equivalent to a physical scale of $\ell \lsim 8-9$ kpc.  Fourth, of the 15 sources in the sample, ten SMGs have radio, mid-IR, and near-IR properties consistent with a higher average/median redshift than radio-selected samples \citep[$z\gsim 3-4$ vs. $z\sim 2.5$ for radio-detected samples: see][]{chapman2005}.  The existence of such a population of high-redshift, hyperluminous starbursts has important consequences for models of galaxy formation, which struggle to account for such extreme objects a $z\gsim 4$.  

\acknowledgements

We thank the referee for their comments, and Lars Hernquist, Doug Finkbeiner, John Huchra, and Dave Sanders for helpful discussions.  The Submillimeter Array is a joint project between the Smithsonian Astrophysical Observatory and the Academia Sinica Institute of Astronomy and Astrophysics and is funded by the Smithsonian Institution and the Academia Sinica.  This work is based on observations made with the {\it Spitzer} Space Telescope, which is operated by the Jet Propulsion Laboratory, California Institute of Technology, under NASA contact 1407.  This work is partially funded by NSF Grant \#0540852.  The JCMT/AzTEC Survey and SK were supported in part by the Korea Science and Engineering Foundation (KOSEF) under a cooperative agreement with the Astrophysical Research Center of the Structure and Evolution of the Cosmos (ARCSEC).  JDY acknowledges support from NASA through Hubble Fellowship grant \#HF-51266.01 awarded by the Space Telescope Science Institute, which is operated by the Association of Universities for Research in Astronomy, Inc., for NASA, under contract NAS 5-26555.   DHH and IA are supported in part by CONACyT.

{\em Facilities:} \facility{SMA}, \facility{JCMT}, \facility{Spitzer (IRAC, MIPS)}, \facility{HST (ACS)}, \facility{Subaru (Suprime-Cam)}, \facility{VLA}

\bibliographystyle{apj}
\bibliography{../../smg}

\begin{thebibliography}{143}
\expandafter\ifx\csname natexlab\endcsname\relax\def\natexlab#1{#1}\fi

\bibitem[{{Appleton} {et~al.}(2004)}]{appleton2004}
{Appleton}, P.~N. {et~al.} 2004, \apjs, 154, 147

\bibitem[{{Arendt} {et~al.}(1998)}]{aredt1998}
{Arendt}, R.~G. {et~al.} 1998, \apj, 508, 74

\bibitem[{{Ashby} {et~al.}(2006)}]{ashby2006}
{Ashby}, M.~L.~N. {et~al.} 2006, \apj, 644, 778

\bibitem[{{Austermann} {et~al.}(2009)}]{austermann2009}
{Austermann}, J.~E. {et~al.} 2009, ArXiv e-prints

\bibitem[{{Barger} {et~al.}(2000){Barger}, {Cowie}, \& {Richards}}]{barger2000}
{Barger}, A.~J., {Cowie}, L.~L., \& {Richards}, E.~A. 2000, \aj, 119, 2092

\bibitem[{{Barger} {et~al.}(1998)}]{barger1998}
{Barger}, A.~J. {et~al.} 1998, \nat, 394, 248

\bibitem[{{Barmby} {et~al.}(2006)}]{barmby2006}
{Barmby}, P. {et~al.} 2006, \apj, 642, 126

\bibitem[{{Baugh} {et~al.}(2005){Baugh}, {Lacey}, {Frenk}, {Granato}, {Silva},
  {Bressan}, {Benson}, \& {Cole}}]{baugh2005}
{Baugh}, C.~M., {Lacey}, C.~G., {Frenk}, C.~S., {Granato}, G.~L., {Silva}, L.,
  {Bressan}, A., {Benson}, A.~J., \& {Cole}, S. 2005, \mnras, 356, 1191

\bibitem[{{Beasley} {et~al.}(2002){Beasley}, {Gordon}, {Peck}, {Petrov},
  {MacMillan}, {Fomalont}, \& {Ma}}]{beasley2002}
{Beasley}, A.~J., {Gordon}, D., {Peck}, A.~B., {Petrov}, L., {MacMillan},
  D.~S., {Fomalont}, E.~B., \& {Ma}, C. 2002, \apjs, 141, 13

\bibitem[{{Bertoldi} {et~al.}(2007)}]{bertoldi2007}
{Bertoldi}, F. {et~al.} 2007, \apjs, 172, 132

\bibitem[{{Biggs} \& {Ivison}(2008)}]{biggs2008}
{Biggs}, A.~D. \& {Ivison}, R.~J. 2008, \mnras, 385, 893

\bibitem[{{Blain} {et~al.}(2004){Blain}, {Chapman}, {Smail}, \&
  {Ivison}}]{blain2004}
{Blain}, A.~W., {Chapman}, S.~C., {Smail}, I., \& {Ivison}, R. 2004, \apj, 611,
  725

\bibitem[{{Blain} \& {Longair}(1993)}]{blain1993}
{Blain}, A.~W. \& {Longair}, M.~S. 1993, \mnras, 264, 509

\bibitem[{{Blain} {et~al.}(1999){Blain}, {Smail}, {Ivison}, \&
  {Kneib}}]{blain1999}
{Blain}, A.~W., {Smail}, I., {Ivison}, R.~J., \& {Kneib}, J.-P. 1999, \mnras,
  302, 632

\bibitem[{{Blain} {et~al.}(2002){Blain}, {Smail}, {Ivison}, {Kneib}, \&
  {Frayer}}]{blain2002}
{Blain}, A.~W., {Smail}, I., {Ivison}, R.~J., {Kneib}, J.-P., \& {Frayer},
  D.~T. 2002, \physrep, 369, 111

\bibitem[{{Bondi} {et~al.}(2008){Bondi}, {Ciliegi}, {Schinnerer}, {Smol{\v
  c}i{\'c}}, {Jahnke}, {Carilli}, \& {Zamorani}}]{bondi2008}
{Bondi}, M., {Ciliegi}, P., {Schinnerer}, E., {Smol{\v c}i{\'c}}, V., {Jahnke},
  K., {Carilli}, C., \& {Zamorani}, G. 2008, \apj, 681, 1129

\bibitem[{{Borys} {et~al.}(2005){Borys}, {Smail}, {Chapman}, {Blain},
  {Alexander}, \& {Ivison}}]{borys2005}
{Borys}, C., {Smail}, I., {Chapman}, S.~C., {Blain}, A.~W., {Alexander}, D.~M.,
  \& {Ivison}, R.~J. 2005, \apj, 635, 853

\bibitem[{{Boyle} {et~al.}(2007)}]{boyle2007}
{Boyle}, B.~J. {et~al.} 2007, \mnras, 376, 1182

\bibitem[{{Browne} {et~al.}(1998){Browne}, {Wilkinson}, {Patnaik}, \&
  {Wrobel}}]{browne1998}
{Browne}, I.~W.~A., {Wilkinson}, P.~N., {Patnaik}, A.~R., \& {Wrobel}, J.~M.
  1998, \mnras, 293, 257

\bibitem[{{Capak} {et~al.}(2007)}]{capak2007}
{Capak}, P. {et~al.} 2007, \apjs, 172, 99

\bibitem[{{Capak} {et~al.}(2008)}]{capak2008}
---. 2008, \apjl, 681, L53

\bibitem[{{Carilli} \& {Yun}(1999)}]{carilli1999}
{Carilli}, C.~L. \& {Yun}, M.~S. 1999, \apjl, 513, L13

\bibitem[{{Casey} {et~al.}(2009)}]{casey2009}
{Casey}, C.~M. {et~al.} 2009, MNRAS, in press [astro-ph/0906.5346]

\bibitem[{{Chapman} {et~al.}(2003){Chapman}, {Blain}, {Ivison}, \&
  {Smail}}]{chapman2003a}
{Chapman}, S.~C., {Blain}, A.~W., {Ivison}, R.~J., \& {Smail}, I.~R. 2003,
  \nat, 422, 695

\bibitem[{{Chapman} {et~al.}(2005){Chapman}, {Blain}, {Smail}, \&
  {Ivison}}]{chapman2005}
{Chapman}, S.~C., {Blain}, A.~W., {Smail}, I., \& {Ivison}, R.~J. 2005, \apj,
  622, 772

\bibitem[{{Chapman} {et~al.}(2004){Chapman}, {Smail}, {Windhorst}, {Muxlow}, \&
  {Ivison}}]{chapman2004}
{Chapman}, S.~C., {Smail}, I., {Windhorst}, R., {Muxlow}, T., \& {Ivison},
  R.~J. 2004, \apj, 611, 732

\bibitem[{{Condon}(1992)}]{condon1992}
{Condon}, J.~J. 1992, \araa, 30, 575

\bibitem[{{Condon}(1997)}]{condon1997}
---. 1997, \pasp, 109, 166

\bibitem[{{Coppin} {et~al.}(2006)}]{coppin2006}
{Coppin}, K. {et~al.} 2006, \mnras, 372, 1621

\bibitem[{{Coppin} {et~al.}(2008{\natexlab{a}})}]{coppin2008.sharcii}
---. 2008{\natexlab{a}}, \mnras, 384, 1597

\bibitem[{{Coppin} {et~al.}(2009)}]{coppin2009}
---. 2009, MNRAS, in press [astro-ph/0902.4464]

\bibitem[{{Coppin} {et~al.}(2008{\natexlab{b}})}]{coppin2008.submmqso}
{Coppin}, K.~E.~K. {et~al.} 2008{\natexlab{b}}, \mnras, 389, 45

\bibitem[{{Cowie} {et~al.}(2002){Cowie}, {Barger}, \& {Kneib}}]{cowie2002}
{Cowie}, L.~L., {Barger}, A.~J., \& {Kneib}, J.-P. 2002, \aj, 123, 2197

\bibitem[{{Cowie} {et~al.}(2009){Cowie}, {Barger}, {Wang}, \&
  {Williams}}]{cowie2009}
{Cowie}, L.~L., {Barger}, A.~J., {Wang}, W.-H., \& {Williams}, J.~P. 2009,
  \apjl, 697, L122

\bibitem[{{Daddi} {et~al.}(2009{\natexlab{a}}){Daddi}, {Dannerbauer}, {Krips},
  {Walter}, {Dickinson}, {Elbaz}, \& {Morrison}}]{daddi2009b}
{Daddi}, E., {Dannerbauer}, H., {Krips}, M., {Walter}, F., {Dickinson}, M.,
  {Elbaz}, D., \& {Morrison}, G.~E. 2009{\natexlab{a}}, ApJL, in press
  [astro-ph/0903.3046]

\bibitem[{{Daddi} {et~al.}(2009{\natexlab{b}})}]{daddi2008}
{Daddi}, E. {et~al.} 2009{\natexlab{b}}, ApJ, in press [astro-ph/0810.3108]

\bibitem[{{Dale} \& {Helou}(2002)}]{dale2002}
{Dale}, D.~A. \& {Helou}, G. 2002, \apj, 576, 159

\bibitem[{{Dannerbauer} {et~al.}(2008){Dannerbauer}, {Walter}, \&
  {Morrison}}]{dannerbauer2008}
{Dannerbauer}, H., {Walter}, F., \& {Morrison}, G. 2008, \apjl, 673, L127

\bibitem[{{Dannerbauer} {et~al.}(2002)}]{dannerbauer2002}
{Dannerbauer}, H. {et~al.} 2002, \apj, 573, 473

\bibitem[{{Dannerbauer} {et~al.}(2004)}]{dannerbauer2004}
---. 2004, \apj, 606, 664

\bibitem[{{Devlin} {et~al.}(2009)}]{devlin2009}
{Devlin}, M.~J. {et~al.} 2009, \nat, 458, 737

\bibitem[{{Donley} {et~al.}(2007){Donley}, {Rieke}, {P{\'e}rez-Gonz{\'a}lez},
  {Rigby}, \& {Alonso-Herrero}}]{donley2007}
{Donley}, J.~L., {Rieke}, G.~H., {P{\'e}rez-Gonz{\'a}lez}, P.~G., {Rigby},
  J.~R., \& {Alonso-Herrero}, A. 2007, \apj, 660, 167

\bibitem[{{Downes} {et~al.}(1986){Downes}, {Peacock}, {Savage}, \&
  {Carrie}}]{downes1986}
{Downes}, A.~J.~B., {Peacock}, J.~A., {Savage}, A., \& {Carrie}, D.~R. 1986,
  \mnras, 218, 31

\bibitem[{{Downes} \& {Solomon}(2003)}]{downes2003}
{Downes}, D. \& {Solomon}, P.~M. 2003, \apj, 582, 37

\bibitem[{{Downes} {et~al.}(1999)}]{downes1999}
{Downes}, D. {et~al.} 1999, \aap, 347, 809

\bibitem[{{Dunlop}(2001)}]{dunlop2001}
{Dunlop}, J.~S. 2001, New Astronomy Reviews, 45, 609

\bibitem[{{Dunlop} {et~al.}(1989){Dunlop}, {Peacock}, {Savage}, {Lilly},
  {Heasley}, \& {Simon}}]{dunlop1989}
{Dunlop}, J.~S., {Peacock}, J.~A., {Savage}, A., {Lilly}, S.~J., {Heasley},
  J.~N., \& {Simon}, A.~J.~B. 1989, \mnras, 238, 1171

\bibitem[{{Dunne} {et~al.}(2003){Dunne}, {Eales}, {Ivison}, {Morgan}, \&
  {Edmunds}}]{dunne2003}
{Dunne}, L., {Eales}, S., {Ivison}, R., {Morgan}, H., \& {Edmunds}, M. 2003,
  \nat, 424, 285

\bibitem[{{Dwek} {et~al.}(1998)}]{dwek1998}
{Dwek}, E. {et~al.} 1998, \apj, 508, 106

\bibitem[{{Eales} {et~al.}(1999)}]{eales1999}
{Eales}, S. {et~al.} 1999, \apj, 515, 518

\bibitem[{{Eales} {et~al.}(2000)}]{eales2000}
---. 2000, \aj, 120, 2244

\bibitem[{{Efstathiou} \& {Rowan-Robinson}(2003)}]{efstathiou2003}
{Efstathiou}, A. \& {Rowan-Robinson}, M. 2003, \mnras, 343, 322

\bibitem[{{Fazio} {et~al.}(2004)}]{fazio2004}
{Fazio}, G.~G. {et~al.} 2004, \apjs, 154, 10

\bibitem[{{Fixsen} {et~al.}(1998)}]{fixsen1998}
{Fixsen}, D.~J. {et~al.} 1998, \apj, 508, 123

\bibitem[{{Ford} {et~al.}(1998)}]{ford1998}
{Ford}, H.~C. {et~al.} 1998, in Proc. SPIE Vol. 3356, p. 234-248, Space
  Telescopes and Instruments V, Pierre Y. Bely; James B. Breckinridge; Eds.,
  ed. P.~Y. {Bely} \& J.~B. {Breckinridge}, 234--248

\bibitem[{{Frayer} {et~al.}(2000){Frayer}, {Smail}, {Ivison}, \&
  {Scoville}}]{frayer2000}
{Frayer}, D.~T., {Smail}, I., {Ivison}, R.~J., \& {Scoville}, N.~Z. 2000, \aj,
  120, 1668

\bibitem[{{Garrett}(2002)}]{garrett2002}
{Garrett}, M.~A. 2002, \aap, 384, L19

\bibitem[{{Gehrz}(1989)}]{gehrz1989}
{Gehrz}, R. 1989, in IAU Symposium, Vol. 135, Interstellar Dust, ed. L.~J.
  {Allamandola} \& A.~G.~G.~M. {Tielens}, 445

\bibitem[{{Genzel} {et~al.}(2003)}]{genzel2003}
{Genzel}, R. {et~al.} 2003, \apj, 584, 633

\bibitem[{{Georgakakis} {et~al.}(2000){Georgakakis}, {Mobasher}, {Cram},
  {Hopkins}, \& {Rowan-Robinson}}]{georgakakis2000}
{Georgakakis}, A., {Mobasher}, B., {Cram}, L., {Hopkins}, A., \&
  {Rowan-Robinson}, M. 2000, \aaps, 141, 89

\bibitem[{{Greve} {et~al.}(2004)}]{greve2004}
{Greve}, T.~R. {et~al.} 2004, \mnras, 354, 779

\bibitem[{{Greve} {et~al.}(2005)}]{greve2005}
---. 2005, \mnras, 359, 1165

\bibitem[{{Gruppioni} {et~al.}(2003)}]{gruppioni2003}
{Gruppioni}, C. {et~al.} 2003, \mnras, 341, L1

\bibitem[{{Hauser} {et~al.}(1998)}]{hauser1998}
{Hauser}, M.~G. {et~al.} 1998, \apj, 508, 25

\bibitem[{{Hickox} {et~al.}(2007)}]{hickox2007}
{Hickox}, R.~C. {et~al.} 2007, \apj, 671, 1365

\bibitem[{{Ho} {et~al.}(2004){Ho}, {Moran}, \& {Lo}}]{ho2004}
{Ho}, P.~T.~P., {Moran}, J.~M., \& {Lo}, K.~Y. 2004, \apjl, 616, L1

\bibitem[{{Holland} {et~al.}(1999)}]{holland1999}
{Holland}, W.~S. {et~al.} 1999, \mnras, 303, 659

\bibitem[{{Hopkins}(2004)}]{hopkins2004}
{Hopkins}, A.~M. 2004, \apj, 615, 209

\bibitem[{{Hopkins} {et~al.}(2003){Hopkins}, {Afonso}, {Chan}, {Cram},
  {Georgakakis}, \& {Mobasher}}]{hopkins2003}
{Hopkins}, A.~M., {Afonso}, J., {Chan}, B., {Cram}, L.~E., {Georgakakis}, A.,
  \& {Mobasher}, B. 2003, \aj, 125, 465

\bibitem[{{Hopkins} {et~al.}(2008{\natexlab{a}}){Hopkins}, {Cox}, {Kere{\v s}},
  \& {Hernquist}}]{hopkins2007b}
{Hopkins}, P.~F., {Cox}, T.~J., {Kere{\v s}}, D., \& {Hernquist}, L.
  2008{\natexlab{a}}, \apjs, 175, 390

\bibitem[{{Hopkins} {et~al.}(2008{\natexlab{b}}){Hopkins}, {Hernquist}, {Cox},
  \& {Kere{\v s}}}]{hopkins2007a}
{Hopkins}, P.~F., {Hernquist}, L., {Cox}, T.~J., \& {Kere{\v s}}, D.
  2008{\natexlab{b}}, \apjs, 175, 356

\bibitem[{{Hopkins} {et~al.}(2006)}]{hopkins2006}
{Hopkins}, P.~F. {et~al.} 2006, \apjs, 163, 1

\bibitem[{{Huang} {et~al.}(2004)}]{huang2004}
{Huang}, J.-S. {et~al.} 2004, \apjs, 154, 44

\bibitem[{{Hughes} {et~al.}(1998)}]{hughes1998}
{Hughes}, D.~H. {et~al.} 1998, \nat, 394, 241

\bibitem[{{Ibar} {et~al.}(2008)}]{ibar2008}
{Ibar}, E. {et~al.} 2008, \mnras, 386, 953

\bibitem[{{Iono} {et~al.}(2006)}]{iono2006}
{Iono}, D. {et~al.} 2006, \apjl, 640, L1

\bibitem[{{Ivison} {et~al.}(2008){Ivison}, {Morrison}, {Biggs}, {Smail},
  {Willner}, {Gurwell}, {Greve}, {Stevens}, \& {Ashby}}]{ivison2008.agnsb}
{Ivison}, R.~J., {Morrison}, G.~E., {Biggs}, A.~D., {Smail}, I., {Willner},
  S.~P., {Gurwell}, M.~A., {Greve}, T.~R., {Stevens}, J.~A., \& {Ashby},
  M.~L.~N. 2008, \mnras, 390, 1117

\bibitem[{{Ivison} {et~al.}(2002)}]{ivison2002}
{Ivison}, R.~J. {et~al.} 2002, \mnras, 337, 1

\bibitem[{{Ivison} {et~al.}(2004)}]{ivison2004}
---. 2004, \apjs, 154, 124

\bibitem[{{Ivison} {et~al.}(2007)}]{ivison2007}
---. 2007, \mnras, 380, 199

\bibitem[{{Kaviani} {et~al.}(2003){Kaviani}, {Haehnelt}, \&
  {Kauffmann}}]{kaviani2003}
{Kaviani}, A., {Haehnelt}, M.~G., \& {Kauffmann}, G. 2003, \mnras, 340, 739

\bibitem[{{Kelsall} {et~al.}(1998)}]{kelsall1998}
{Kelsall}, T. {et~al.} 1998, \apj, 508, 44

\bibitem[{{Kneib} {et~al.}(2005)}]{kneib2005}
{Kneib}, J.-P. {et~al.} 2005, \aap, 434, 819

\bibitem[{{Koekemoer} {et~al.}(2007)}]{koekemoer2007}
{Koekemoer}, A.~M. {et~al.} 2007, \apjs, 172, 196

\bibitem[{{Kov{\'a}cs} {et~al.}(2006)}]{kovacs2006}
{Kov{\'a}cs}, A. {et~al.} 2006, \apj, 650, 592

\bibitem[{{Lacy} {et~al.}(2004)}]{lacy2004}
{Lacy}, M. {et~al.} 2004, \apjs, 154, 166

\bibitem[{{Laurent} {et~al.}(2005)}]{laurent2005}
{Laurent}, G.~T. {et~al.} 2005, \apj, 623, 742

\bibitem[{{Le Floc'h} {et~al.}(2005)}]{lefloch2005}
{Le Floc'h}, E. {et~al.} 2005, \apj, 632, 169

\bibitem[{{Lilly} {et~al.}(1999){Lilly}, {Eales}, {Gear}, {Hammer}, {Le
  F{\`e}vre}, {Crampton}, {Bond}, \& {Dunne}}]{lilly1999}
{Lilly}, S.~J., {Eales}, S.~A., {Gear}, W.~K.~P., {Hammer}, F., {Le F{\`e}vre},
  O., {Crampton}, D., {Bond}, J.~R., \& {Dunne}, L. 1999, \apj, 518, 641

\bibitem[{{Lutz} {et~al.}(1998)}]{lutz1998}
{Lutz}, D. {et~al.} 1998, \apjl, 505, L103

\bibitem[{{Ma} {et~al.}(1998)}]{ma1998}
{Ma}, C. {et~al.} 1998, \aj, 116, 516

\bibitem[{{Madau} {et~al.}(1996)}]{madau1996}
{Madau}, P. {et~al.} 1996, \mnras, 283, 1388

\bibitem[{{Magnelli} {et~al.}(2009){Magnelli}, {Elbaz}, {Chary}, {Dickinson},
  {Le Borgne}, {Frayer}, \& {Willmer}}]{magnelli2009}
{Magnelli}, B., {Elbaz}, D., {Chary}, R.~R., {Dickinson}, M., {Le Borgne}, D.,
  {Frayer}, D.~T., \& {Willmer}, C.~N.~A. 2009, \aap, 496, 57

\bibitem[{{Maloney} \& {Reynolds}(2000)}]{maloney2000}
{Maloney}, P.~R. \& {Reynolds}, C.~S. 2000, \apjl, 545, L23

\bibitem[{{Marchenko}(2006)}]{marchenko2006}
{Marchenko}, S.~V. 2006, in Astronomical Society of the Pacific Conference
  Series, Vol. 353, Stellar Evolution at Low Metallicity: Mass Loss,
  Explosions, Cosmology, ed. H.~J.~G.~L.~M. {Lamers}, N.~{Langer}, T.~{Nugis},
  \& K.~{Annuk}, 299

\bibitem[{{Mobasher} {et~al.}(2007)}]{mobasher2006}
{Mobasher}, B. {et~al.} 2007, \apjs, 172, 117

\bibitem[{{Narayanan} {et~al.}(2009{\natexlab{a}}){Narayanan}, {Cox},
  {Hayward}, {Younger}, \& {Hernquist}}]{narayanan2009b}
{Narayanan}, D., {Cox}, T.~J., {Hayward}, C., {Younger}, J.~D., \& {Hernquist},
  L. 2009{\natexlab{a}}, MNRAS, submitted [astro-ph/0905.2184]

\bibitem[{{Narayanan} {et~al.}(2009{\natexlab{b}}){Narayanan}, {Hayward},
  {Cox}, {Hernquist}, {Jonsson}, {Younger}, \& {Groves}}]{narayanan2009}
{Narayanan}, D., {Hayward}, C.~C., {Cox}, T.~J., {Hernquist}, L., {Jonsson},
  P., {Younger}, J.~D., \& {Groves}, B. 2009{\natexlab{b}}, ApJL, submitted
  [astro-ph/0904.0004]

\bibitem[{{Neri} {et~al.}(2003)}]{neri2003}
{Neri}, R. {et~al.} 2003, \apjl, 597, L113

\bibitem[{{Pascale} {et~al.}(2009)}]{pascale2009}
{Pascale}, E. {et~al.} 2009, ApJ, submitted [astro-ph/0904.1206]

\bibitem[{{Patnaik} {et~al.}(1992){Patnaik}, {Browne}, {Wilkinson}, \&
  {Wrobel}}]{patnaik1992}
{Patnaik}, A.~R., {Browne}, I.~W.~A., {Wilkinson}, P.~N., \& {Wrobel}, J.~M.
  1992, \mnras, 254, 655

\bibitem[{{Pei} {et~al.}(1999){Pei}, {Fall}, \& {Hauser}}]{pei1999}
{Pei}, Y.~C., {Fall}, S.~M., \& {Hauser}, M.~G. 1999, \apj, 522, 604

\bibitem[{{Perera} {et~al.}(2008)}]{perera2008}
{Perera}, T.~A. {et~al.} 2008, \mnras, 1261

\bibitem[{{Pope} {et~al.}(2006)}]{pope2006}
{Pope}, A. {et~al.} 2006, \mnras, 370, 1185

\bibitem[{{Richards}(2000)}]{richards2000}
{Richards}, E.~A. 2000, \apj, 533, 611

\bibitem[{{Rieke} {et~al.}(2004){Rieke}, , {et~al.}}]{rieke2004}
{Rieke}, G.~H., , {et~al.} 2004, \apjs, 154, 25

\bibitem[{{Risaliti} {et~al.}(2000){Risaliti}, {Gilli}, {Maiolino}, \&
  {Salvati}}]{risaliti2000}
{Risaliti}, G., {Gilli}, R., {Maiolino}, R., \& {Salvati}, M. 2000, \aap, 357,
  13

\bibitem[{{Salpeter}(1955)}]{salpeter1955}
{Salpeter}, E.~E. 1955, \apj, 121, 161

\bibitem[{{Sanders} \& {Mirabel}(1996)}]{sanders1996}
{Sanders}, D.~B. \& {Mirabel}, I.~F. 1996, \araa, 34, 749

\bibitem[{{Sanders} {et~al.}(1988{\natexlab{a}}){Sanders}, {Soifer}, {Elias},
  {Madore}, {Matthews}, {Neugebauer}, \& {Scoville}}]{sanders1988a}
{Sanders}, D.~B., {Soifer}, B.~T., {Elias}, J.~H., {Madore}, B.~F., {Matthews},
  K., {Neugebauer}, G., \& {Scoville}, N.~Z. 1988{\natexlab{a}}, \apj, 325, 74

\bibitem[{{Sanders} {et~al.}(1988{\natexlab{b}}){Sanders}, {Soifer}, {Elias},
  {Neugebauer}, \& {Matthews}}]{sanders1988b}
{Sanders}, D.~B., {Soifer}, B.~T., {Elias}, J.~H., {Neugebauer}, G., \&
  {Matthews}, K. 1988{\natexlab{b}}, \apjl, 328, L35

\bibitem[{{Sanders} {et~al.}(2007)}]{sanders2007}
{Sanders}, D.~B. {et~al.} 2007, \apjs, 172, 86

\bibitem[{{Sault} {et~al.}(1995){Sault}, {Teuben}, \& {Wright}}]{sault1995}
{Sault}, R.~J., {Teuben}, P.~J., \& {Wright}, M.~C.~H. 1995, in ASP Conf. Ser.
  77: Astronomical Data Analysis Software and Systems IV, ed. R.~A. {Shaw},
  H.~E. {Payne}, \& J.~J.~E. {Hayes}, 433

\bibitem[{{Schinnerer} {et~al.}(2007)}]{schinnerer2007}
{Schinnerer}, E. {et~al.} 2007, \apjs, 172, 46

\bibitem[{{Schinnerer} {et~al.}(2008)}]{schinnerer2008}
---. 2008, ArXiv e-prints

\bibitem[{{Scott} \& {Tout}(1989)}]{scott1989}
{Scott}, D. \& {Tout}, C.~A. 1989, \mnras, 241, 109

\bibitem[{{Scott} {et~al.}(2006)}]{dscott2006}
{Scott}, D. {et~al.} 2006, in Bulletin of the American Astronomical Society,
  Vol.~38, Bulletin of the American Astronomical Society, 1072

\bibitem[{{Scott} {et~al.}(2008)}]{scott2008}
{Scott}, K.~S. {et~al.} 2008, \mnras, 385, 2225

\bibitem[{{Scott} {et~al.}(2002)}]{scott2002}
{Scott}, S.~E. {et~al.} 2002, \mnras, 331, 817

\bibitem[{{Scoville} {et~al.}(2007)}]{scoville2007}
{Scoville}, N. {et~al.} 2007, \apjs, 172, 1

\bibitem[{{Scoville} {et~al.}(1993){Scoville}, {Carlstrom}, {Chandler},
  {Phillips}, {Scott}, {Tilanus}, \& {Wang}}]{scoville1993}
{Scoville}, N.~Z., {Carlstrom}, J.~E., {Chandler}, C.~J., {Phillips}, J.~A.,
  {Scott}, S.~L., {Tilanus}, R.~P.~J., \& {Wang}, Z. 1993, \pasp, 105, 1482

\bibitem[{{Serjeant} {et~al.}(2003)}]{serjeant2003}
{Serjeant}, S. {et~al.} 2003, \mnras, 344, 887

\bibitem[{{Sheth} {et~al.}(2004)}]{sheth2004}
{Sheth}, K. {et~al.} 2004, \apjl, 614, L5

\bibitem[{{Smail} {et~al.}(1997){Smail}, {Ivison}, \& {Blain}}]{smail1997}
{Smail}, I., {Ivison}, R.~J., \& {Blain}, A.~W. 1997, \apjl, 490, L5

\bibitem[{{Stern} {et~al.}(2005)}]{stern2005}
{Stern}, D. {et~al.} 2005, \apj, 631, 163

\bibitem[{{Swinbank} {et~al.}(2006)}]{swinbank2006}
{Swinbank}, A.~M. {et~al.} 2006, \mnras, 371, 465

\bibitem[{{Swinbank} {et~al.}(2008)}]{swinbank2008}
---. 2008, \mnras, 391, 420

\bibitem[{{Tacconi} {et~al.}(2006)}]{tacconi2006}
{Tacconi}, L.~J. {et~al.} 2006, \apj, 640, 228

\bibitem[{{Tacconi} {et~al.}(2008)}]{tacconi2008}
---. 2008, \apj, 680, 246

\bibitem[{{Taniguchi} {et~al.}(2007)}]{taniguchi2007}
{Taniguchi}, Y. {et~al.} 2007, \apjs, 172, 9

\bibitem[{{Viero} {et~al.}(2009)}]{viero2009}
{Viero}, M.~P. {et~al.} 2009, ApJ, submitted [astro-ph/0904.1200]

\bibitem[{{Wang} {et~al.}(2009){Wang}, {Barger}, \& {Cowie}}]{wang2008}
{Wang}, W.-H., {Barger}, A.~J., \& {Cowie}, L.~L. 2009, \apj, 690, 319

\bibitem[{{Wang} {et~al.}(2004){Wang}, {Cowie}, \& {Barger}}]{wang2004}
{Wang}, W.-H., {Cowie}, L.~L., \& {Barger}, A.~J. 2004, \apj, 613, 655

\bibitem[{{Wang} {et~al.}(2007)}]{wang2007}
{Wang}, W.-H. {et~al.} 2007, \apjl, 670, L89

\bibitem[{{Webb} {et~al.}(2003)}]{webb2003}
{Webb}, T.~M. {et~al.} 2003, \apj, 587, 41

\bibitem[{{Wilson} {et~al.}(2008)}]{wilson2008b}
{Wilson}, G.~W. {et~al.} 2008, \mnras, 390, 1061

\bibitem[{{Younger} {et~al.}(2009{\natexlab{a}}){Younger}, {Hayward},
  {Narayanan}, {Cox}, {Hernquist}, \& {Jonsson}}]{younger2009.warmcold}
{Younger}, J.~D., {Hayward}, C.~C., {Narayanan}, D., {Cox}, T.~J., {Hernquist},
  L., \& {Jonsson}, P. 2009{\natexlab{a}}, \mnras, 396, L66

\bibitem[{{Younger} {et~al.}(2009{\natexlab{b}}){Younger}, {Omont}, {Fiolet},
  {Huang}, {Fazio}, {Lai}, {Polletta}, {Rigopoulou}, \&
  {Zylka}}]{younger2008.egsulirgs}
{Younger}, J.~D., {Omont}, A., {Fiolet}, N., {Huang}, J.-S., {Fazio}, G.~G.,
  {Lai}, K., {Polletta}, M., {Rigopoulou}, D., \& {Zylka}, R.
  2009{\natexlab{b}}, \mnras, 394, 1685

\bibitem[{{Younger} {et~al.}(2007)}]{younger2007}
{Younger}, J.~D. {et~al.} 2007, \apj, 671, 1531

\bibitem[{{Younger} {et~al.}(2008{\natexlab{a}})}]{younger2008}
---. 2008{\natexlab{a}}, \mnras, 387, 707

\bibitem[{{Younger} {et~al.}(2008{\natexlab{b}})}]{younger2008highres}
---. 2008{\natexlab{b}}, \apj, 688, 59

\bibitem[{{Yun} \& {Carilli}(2002)}]{yun2002}
{Yun}, M.~S. \& {Carilli}, C.~L. 2002, \apj, 568, 88

\bibitem[{{Yun} {et~al.}(2008)}]{yun2008}
{Yun}, M.~S. {et~al.} 2008, \mnras, 389, 333

\end{thebibliography}

\end{document}